\documentclass[aps,prx,longbibliography,10pt]{revtex4-2}
\usepackage{graphicx} 
\usepackage[braket, qm]{qcircuit} 
\usepackage{amsmath}
\usepackage{amssymb}
\usepackage{dsfont}
\usepackage[colorlinks=true, allcolors=blue]{hyperref}
\usepackage[dvipsnames]{xcolor}
\usepackage{array}
\usepackage{amsthm}
\usepackage[ruled,vlined,linesnumbered]{algorithm2e}
\usepackage{microtype}
\usepackage{thmtools}
\usepackage{thm-restate}
\usepackage{braket}
\usepackage{geometry}
\usepackage{tikz}
\usepackage{booktabs}
\usepackage{mathtools}
\usepackage{orcidlink}
\usepackage{placeins}   
\usepackage{makecell}
\usepackage{rotating}
\usepackage{orcidlink}
\usepackage[capitalise,nameinlink]{cleveref}
\usepackage{verbatim}
\usepackage{tikz}

\usepackage[dvipsnames]{xcolor}

\DeclareMathOperator*{\mux}{\mathsf{mux}}

\begin{document}

\title{Efficient Block Encodings for Ensembles of Molecular Hamiltonians}

\newcommand{\BI}{\affiliation{
Quantum Lab, Boehringer Ingelheim, 55218 Ingelheim am Rhein, Germany}}

\newcommand{\BIMedChem}{\affiliation{Medicinal Chemistry, Boehringer Ingelheim Pharma GmbH \& Co. KG, \\88397 Biberach an der Riß, Germany}}

\newcommand{\PSIQ}{\affiliation{PsiQuantum, 700 Hansen Way, Palo Alto, CA 94304, USA}}

\author{Cristian L.~Cortes}
\email{ccortes@psiquantum.com}
\PSIQ

\author{Andrew J.~Jenkins}
\PSIQ

\author{William Pol}
\PSIQ

\author{Dario Rocca}
\PSIQ

\author{Mark Steudtner}
\PSIQ

\author{Gian-Luca R.~Anselmetti\orcidlink{0000-0002-8073-3567}}
\BI

\author{Matthias Degroote\orcidlink{0000-0002-8850-7708}}
\email{matthias.degroote@boehringer-ingelheim.com}
\BI

\author{Nikolaj Moll\orcidlink{0000-0001-5645-4667}}
\BI

\author{Raffaele Santagati\orcidlink{0000-0001-9645-0580}}
\BI

\author{Michael Streif\orcidlink{0000-0002-7509-4748}}
\BI

\author{Markus R. Hermann}
\BIMedChem

\begin{abstract}
    Accurately characterizing molecular properties and reactivity often requires electronic-structure calculations across an ensemble of molecular geometries. Quantum algorithms that process such ensembles coherently could address this challenge, but they require Hamiltonian block encoding circuits that combine advanced factorization techniques with sublinear scaling in ensemble size. Here, we introduce SVD-THC, which extends tensor hypercontraction to sets of Hamiltonians defined at different nuclear geometries through a singular value decomposition of the ensemble. We use SVD-THC to construct a single multiplexed block encoding circuit for all geometries in an ensemble. We then evaluate the approach on ensembles generated by varying internal rotational coordinates in 1-bromo-2-chloroethane as well as in glycine and alanine dipeptides, corresponding to established model systems for peptide conformational landscapes, with up to 5,184 geometries. Across several resolutions of the internal-coordinate grids, the Toffoli complexity of the resulting circuits grows only weakly with the number of geometries, changing by at most 12\% over grids that differ in size by up to 576$\times$. In an energy-thresholding circuit, this yields 35--200$\times$ fewer Toffoli gates than slicing together single-geometry block encodings. This approach could help make coherent processing of molecular geometry ensembles a useful primitive for fault-tolerant quantum computing.
\end{abstract}
\maketitle

\section{Introduction}
Ground state energy estimation is often described as the central use case for quantum computing in chemistry~\cite{Reiher2017, Cao2019, Bauer2020QuantumScience, McArdle2020QuantumChemistry,Dalzell2025,Babbush2026_Applications}. Indeed, a large body of work has been dedicated to the solution of the electronic structure problem, where the quantum computer would find the ground state of an electronic Hamiltonian for a fixed set of atomic coordinates. However, this static image of quantum chemistry rarely delivers answers to pharmaceutically relevant questions~\cite{Santagati2024,Zhou2026_DrugDisc}. Many interesting properties can instead be derived from the shape of the potential energy surface (PES) -- the energy functional across atomic geometries: In geometry optimization, one searches for the minimal point in the PES which is where the molecule is most stable. During a molecular dynamics (MD) run, one follows the nuclei as their positions evolve over time which naturally generates the low-energy conformations or allows one to follow reagents as they chemically transform into products. The slope of the PES indicates the direction in which the nuclei experience a force during the time evolution~\cite{Hao2022_GeomOpt, Tautermann2015_MolDynDrugs, Schlegel2011Geometry, Kassal2009_MolPropGeomOpt, Kassal2008_ChemDyn}. Rotatable bonds determine the relative positions of different active parts of small-molecule drugs. The lowest-energy dihedral allows one to understand how stable the bioactive conformation is. Rotational profiles illuminate where barriers to torsional movement exist and which areas of phase space are accessible by thermal motion. That informs the decision whether a certain conformation of a drug molecule is stable and whether other forms are accessible under thermal motion. These ensemble functions provide important data that helps make decisions in the drug discovery cycle~\cite{Perola2004_ConfAnalysLig}.
To learn these properties in quantum computing, one would either have to solve the electronic structure problem for multiple geometries, or encode electronic information of multiple geometries in a state of the form
\begin{align}
\label{eq:pesstate}
|\Psi\rangle = \sum_{\boldsymbol{\theta}} c_{\boldsymbol{\theta}} |\boldsymbol{\theta}\rangle |\psi_{\boldsymbol{\theta},0}\rangle\, ,
\end{align}
where $|\psi_{\boldsymbol{\theta},0}\rangle$ are electronic ground states of the geometry vector $\boldsymbol{\theta}$ encoded in some index state $|\boldsymbol{\theta}\rangle$~\cite{Gilyen2017,OBrien2022_Forces, steudtner2023fault}. With such a state, one could aggregate information across multiple geometries. The stretch goal would be to eventually encode a version of the PES itself, where the coefficients $c_{\boldsymbol{\theta}}$ become a function of the ground state energies~\cite{Kosugi2022_OptMolGeom, Kassal2008_ChemDyn}. 
For the preparation of a PES state in Eq.~\eqref{eq:pesstate}, as well as operations such as free energy calculation and geometry optimization, one needs coherent and efficient access to electronic structure Hamiltonians of all geometries~\cite{Huang2026fullqubit,Simon2024Improved}. That is, we need to efficiently encode Hamiltonians that are a functional of the atomic coordinates. 

To be competitive with methods relying on repeated energy estimation and classical postprocessing for each geometry, one must aim for a compression sublinear in the number of geometries considered. For a list of $M$ geometries $\boldsymbol{\theta}$, the total cost of encoding the geometry-dependent Hamiltonian $H(\boldsymbol{\theta})$
must scale as $o(M)$ times the average cost of encoding a single-geometry Hamiltonian.
Since the cost of encoding single-geometry Hamiltonians has gradually been reduced through years of intense research, a tight baseline is set: by replacing time evolution with a quantum walk operator, qubitization has greatly accelerated Hamiltonian simulation through block encodings~\cite{low2019hamiltonian,poulin2018quantum, berry2018improved, Gilyen2019_QSVT, Low2017_QSP}. Factorization algorithms such as single factorization, sparse factorization~\cite{berry2019qubitization}, double factorization (DF)~\cite{von2021quantum,Motta2021LowRank}, tensor hypercontraction (THC)~\cite{Hohenstein2012_THC, lee2021even} as well as double-factorized tensor hypercontraction (DFTHC)~\cite{low2025fast} exploit patterns in the Hamiltonians to compress block encoding circuits and speed up the quantum algorithm. Recently, the block-invariant symmetry shift~\cite{loaiza2023block,caesura2025faster} has been used to push this compression towards its theoretical limit~\cite{Cortes2024Assessing}. To be competitive with repeated queries to single-geometry energy estimation, Hamiltonian simulation for geometric ensembles must draw inspiration from the same tools. In Hamiltonian factorization techniques, the strategy is to reduce the amount of redundant information loaded onto the quantum computer and instead generate the data procedurally from a compressed representation obtained with classical preprocessing. The same strategy could be applied to geometric ensembles. Relevant ensembles of Hamiltonians often share a large similarity and differ only in a few coordinates. It is common that the atomic composition and even the relative positions of atoms in ligands further removed from the active center remain the same. A set of Hamiltonians generated over a grid of coordinates may consequently have a lot of redundant information. This gives us confidence that we can devise a Hamiltonian loading scheme that makes use of this extra degree of redundancy and that is capable of constructing a coherent block encoding of the full ensemble that is cheaper than loading each geometry sequentially.

In this paper, we develop SVD-THC, a compression technique that fits an ensemble of $M$ electronic structure Hamiltonians into the same block encoding. This technique allows us to time-evolve an ensemble of various geometries, encoded into a quantum state, as a whole. In the pursuit of an algorithmic complexity sublinear in $M$, we propose a singular value decomposition (SVD) of the ensemble, that expresses the electronic interactions of all geometries as a superposition of fewer than $M$, $\boldsymbol{\theta}$-independent Hamiltonians. All these SVD Hamiltonians are compressed with THC and BLISS techniques, which calls for a multiplexed THC circuit that is efficient for a set of Hamiltonians with individual THC ranks.

We apply our SVD-THC compression technique to the Hamiltonian ensembles of 1-bromo-2-chloroethane (BCE), glycine dipeptide and alanine dipeptide. These model systems increase the hardness of modeling from a one-dimensional PES to a challenging two-dimensional benchmark for machine-learned interatomic potentials (MLIPs)~\cite{Fedik2025Challenges}. For each of these three molecular systems we perform the numerical decomposition of the Hamiltonian ensemble and investigate potential energy surfaces. Quantum resource estimates (QREs) are then obtained in the form of block encoding 1-norms, Toffoli gate counts and active volume (AV) for the energy thresholding circuit, which isolates the low-energy subspace of an ensemble. We discuss the compression factors achieved through SVD-THC, by comparing the thresholding circuit to its counterpart in the picture of repeated single-geometry THC circuits.

We find that the advantage of SVD-THC grows with the number of geometries. Since a single SVD-THC circuit serves the whole ensemble, its cost stays nearly constant as we refine the grid, whereas repeating single-geometry circuits grows in proportion to the number of geometries. On the largest grids, our approach reduces the Toffoli count by $35$--$41\times$ for BCE with 73 geometries, $144\times$ for glycine dipeptide with 2,664 geometries, and $201\times$ for alanine dipeptide with 5,184 geometries, while keeping the CCSD correlation energy within $0.6$~mHa for all configurations. Even against the most economical alternative, corresponding to a separate phase estimation for each geometry, we obtain a $10$--$60\times$ reduction. Ultimately, we find that the compression starts paying off for grids with tens of geometries, with an advantage that continues to grow as the grid is further refined.

The rest of this paper is organized as follows: first we lay out the problem setting and review BLISS-THC in Section~\ref{sec:setting}, and develop the SVD-THC compression in Section~\ref{sec:svdthc}, before we draw its quantum circuit in Section~\ref{sec:circuits}.
In Section~\ref{sec:results}, we present numerical results for the SVD-THC factorization applied to our benchmark molecules, and in Section~\ref{sec:results2}, we obtain QREs for the thresholding circuit. We close the paper with the conclusions and outlook in Section~\ref{sec:conclusion}.

\section{Problem Setting and Background}
\label{sec:setting}
In this section, we define the problem setting by describing ensembles of geometry-dependent Hamiltonians mathematically. Since Hamiltonian factorization is going to have an important role in this work, we briefly review the high-level ideas behind BLISS-THC~\cite{caesura2025faster}.

\subsection{Parameterized electronic structure Hamiltonians}
\label{sec:param}
\noindent
We consider the parameterized electronic structure Hamiltonian in second quantization,
\begin{equation}
H(\boldsymbol{\theta}) =
\mathcal{E}_{\boldsymbol{\theta}}\mathds{1} + \sum_{pq}\sum_{\sigma}
h_{pq}(\boldsymbol{\theta})\,a_{p\sigma}^\dagger a_{q\sigma} +
\frac{1}{2} \sum_{pqrs}\sum_{\sigma\tau} g_{pqrs}(\boldsymbol{\theta})\,
a_{p\sigma}^\dagger a_{r\tau}^\dagger a_{s\tau} a_{q\sigma} ,
\label{eq:ham}
\end{equation}
where $p,q,r,s$ label $N$ spatial orbitals, $\sigma,\tau\in\{\uparrow,\downarrow\}$ denote spin, and $h_{pq}$ and $g_{pqrs}$ are the one- and two-electron integrals in chemist's notation, assumed to be real. The scalar $\mathcal{E}_{\boldsymbol{\theta}}$ contains the nuclear repulsion energy and the mean-field contribution from excluded occupied orbitals. Although this term is typically omitted in block encoding schemes because its contribution can be added during classical postprocessing, it must be retained here since it contributes to the relative energies between configurations. Note that Eq.~\eqref{eq:ham} supplies Hamiltonian integrals as a function of $\boldsymbol{\theta}$; obtaining the full ground-state energy $E(\boldsymbol{\theta})$ still requires solving the electronic structure problem at each geometry.

The parameter vector, $\boldsymbol{\theta}=(\theta_1,\ldots,\theta_d)$, contains $d$ arbitrary coordinates. For example, $\theta_i$ could be a torsion angle, which quantifies the rotation about a chemical bond: for four consecutively bonded atoms A--B--C--D, it is the angle between the plane spanned by A, B, and C and the plane spanned by B, C, and D. If the bond lies along the main chain, or backbone, of a molecule such as a peptide, this angle is called a backbone dihedral. We sample these coordinates on a product grid $\boldsymbol{\theta_n}$, where $\boldsymbol{n}=(n_1,\ldots,n_d),$ with $ n_i\in\{0,\ldots,M_i-1\}.$ The grid therefore contains $ M=\prod_{i=1}^{d}M_i$ total configurations. For quantum circuit constructions in this work, we address each configuration using a single flattened index $\theta\in\{0,\ldots,M-1\}$, and we use $\boldsymbol{\theta_n}$ and $\theta$ interchangeably, with computational states $|\theta\rangle$ indexing geometries in quantum registers.

A naive block encoding of Eq.~\eqref{eq:ham} treats every grid point independently. Its cost therefore scales linearly with $M$ and does not exploit any similarity between neighboring geometries. Our proposed approach instead expands the parameterized Hamiltonian in a data-dependent basis: up to a geometry-dependent energy shift, $H(\boldsymbol{\theta})$ is written as a linear combination of $K$ fixed, geometry-independent operators $H_\ell$ with geometry-dependent coefficients. Each $H_\ell$ has the form of a molecular electronic-structure Hamiltonian, i.e., a sum of one- and two-body fermionic terms, so that established factorization techniques can be applied to it. We focus on BLISS-THC for this manuscript but other common factorizations should be compatible. Note that the construction we present below is agnostic to how the integrals are obtained and applies generally in the all-electron, frozen-core, and active-space settings. We illustrate this explicitly in Section~\ref{sec:results} using the frozen-core picture, where core electrons are projected out and the remaining orbitals are L\"owdin orthogonalized. By contrast, selecting a consistent active space that varies smoothly across the grid is more difficult since active space orbitals can mix or reorder as the geometry changes. We defer this problem to future work.

\subsection{BLISS-THC}
\label{sec:blissthc}

For a specific fermionic Hamiltonian, such as the typical electronic structure Hamiltonian, which corresponds to a single configuration of Eq.~\eqref{eq:ham},
our construction uses the block-invariant symmetry-shifted tensor hypercontraction (BLISS-THC) factorization introduced in Ref.~\cite{caesura2025faster}. For completeness, we summarize the BLISS-THC approach based on the Majorana representation. Expanding the annihilation operator as $a_{p\sigma}= (\hat\gamma_{p\sigma,0}+i\hat\gamma_{p\sigma,1})/2$, the block-invariant electronic structure Hamiltonian may be written as, 
\begin{align}
H_{\mathrm{BI}} = \frac{i}{2}\sum_{pq,\sigma}\kappa^{(\mathrm{BI})}_{pq}
\hat\gamma_{p\sigma,0}\hat\gamma_{q\sigma,1}
- \frac{1}{8}\sum_{pqrs,\sigma\tau} g^{(\mathrm{BI})}_{pqrs}
\hat\gamma_{p\sigma,0}\hat\gamma_{q\sigma,1}
\hat\gamma_{r\tau,0}\hat\gamma_{s\tau,1},
\label{eq:hbi}
\end{align}
with block-invariant coefficients
\begin{align}
\kappa^{(\mathrm{BI})}_{pq} &=
h_{pq} - \tfrac{1}{2}\textstyle\sum_r g_{prrq}
+ \sum_r g^{(\mathrm{BI})}_{pqrr}
- \alpha_1\delta_{pq}
+ \tfrac{1}{2}\beta_{pq}\eta, \label{eq:kappa}\\
g^{(\mathrm{BI})}_{pqrs} &=
g_{pqrs} - \alpha_2\delta_{pq}\delta_{rs}
- \tfrac{1}{2}\big(\beta_{pq}\delta_{rs}
+ \delta_{pq}\beta_{rs}\big). \label{eq:gbi}
\end{align}
Here, the scalars $\alpha_1$ and $\alpha_2$, together with the one-body matrix $\beta_{pq}$, parameterize a shift by a function of the particle number operator $\hat N$ that vanishes identically in the $\eta$-electron sector. Thus, within that sector, $H_{\mathrm{BI}}$ has the same eigenvalue spectrum as $H$, while its 1-norm can become substantially smaller. The block-invariant two-body tensor admits the THC decomposition,
\begin{align}
g^{(\mathrm{BI})}_{pqrs}
= \sum_{\mu,\nu=1}^{R}
\tilde\zeta_{\mu\nu}\,
\chi^{\mu}_{p}\chi^{\mu}_{q}
\chi^{\nu}_{r}\chi^{\nu}_{s},
\label{eq:thc}
\end{align}
where $R$ is the THC rank. The leaf unit vectors $\chi^\mu$ are implemented on the quantum computer using Givens rotations, while the core tensor $\tilde\zeta_{\mu\nu}$ is loaded through state preparation. The resulting block encoding has 1-norm,
\begin{align}
\tilde\lambda_{\mathrm{THC}} = \sum_k |\tilde t_k|
+ \frac{1}{2}\sum_{\mu,\nu} |\tilde\zeta_{\mu\nu}|
- \frac{1}{4}\sum_{\mu} |\tilde\zeta_{\mu\mu}|,
\label{eq:lam_thc}
\end{align}
where $\tilde t_k$ are the eigenvalues of $\kappa^{(\mathrm{BI})}$. Eq.~\eqref{eq:lam_thc} is the objective that BLISS-THC minimizes along with the residual reconstruction error, and it is the per-mode input to everything that follows.

\section{SVD-parameterized BLISS-THC}
\label{sec:svdthc}
The previous section described how BLISS-THC compresses a single Hamiltonian at a fixed geometry. We now extend this construction to an ensemble of Hamiltonians evaluated on a product grid of geometries. The objective here is to represent the full ensemble using a small number of geometry-independent Hamiltonians $H_\ell$, where each Hamiltonian is multiplied by a scalar function of ${\theta}$. Finding such a representation allows us to apply a tensor factorization method, such as BLISS-THC, to each mode separately. We name this technique SVD-THC.

A natural starting point requires obtaining the geometry-dependent functions defined from a fixed basis. Fourier, Chebyshev, and Walsh-Hadamard transforms are possible choices~\cite{Welch_2014}. One of their advantages is that their functions are known ahead of time and are well-behaved. Fourier and Chebyshev bases are effective when the Hamiltonian coefficients vary smoothly across the grid. Walsh-Hadamard bases can be useful when the coefficients have a piecewise or otherwise simple structure. However, these assumptions are not always appropriate for constrained potential-energy-surface scans. In this work, we consider potential energy scans of two kinds. In a rigid scan, only the scanned coordinates change and every other internal coordinate is held fixed. In a relaxed scan, at each point of such a scan, the unconstrained coordinates are geometry optimized resulting in molecular configurations that can change abruptly as the scanned coordinates vary. Relaxed scans give physically relevant energy surfaces, but as we show in Sec.~\ref{sec:results}, they are also harder to compress. Conical intersections can also introduce additional nonanalytic behavior, including cusps. A fixed basis may therefore require many modes to represent these types of features accurately.

In this work, we propose constructing the geometry basis directly from the electronic tensors themselves. The basis should therefore adapt to the variation present in the Hamiltonian, including non-smooth features caused by geometry optimization or conical intersections. This basis is obtained from a singular value decomposition applied to the integral tensors sampled across the grid. For large grids, we compute the relevant singular vectors through an $M\times M$ Gram matrix rather than by forming the much larger orbital-space covariance matrices. This section describes the full construction. We first reshape the integral tensors into snapshot matrices. We then use a single joint geometry basis for the one- and two-body terms. Finally, we show how each retained mode defines a geometry-independent chemistry-like Hamiltonian that can be compressed with BLISS-THC.

\subsection{Snapshot matrices}
\label{sec:snapshots}

We begin by reshaping the one- and two-electron integral tensors sampled across the geometry grid. At every grid point, both tensors are computed in the consistent orbital representation defined in Sec.~\ref{sec:param}. We then reshape each tensor into a \emph{snapshot matrix}, whose rows correspond to geometries and whose columns correspond to flattened orbital indices. For the one-body channel, we use the Majorana one-body coefficient that appears in the electronic structure Hamiltonian,
\begin{align}
\kappa_{pq}({\theta}) =
h_{pq}({\theta})
+ \sum_r g_{pqrr}({\theta})
- \tfrac{1}{2}\sum_r g_{prrq}({\theta}).
\end{align}
We work with \(\kappa_{pq}\), rather than \(h_{pq}\), because \(\kappa_{pq}\) is the one-body matrix that enters the downstream Majorana Hamiltonian representation. The corresponding snapshot matrix is defined as,
\begin{equation}
{A}_{\theta,(pq)}
\Leftrightarrow
\kappa_{pq}({\theta}),
\qquad
{A}\in\mathbb{R}^{M\times N^2}.
\label{eq:snapshot1}
\end{equation}
For the two-body channel, we retain only the entries that are unique under the usual eight-fold permutational symmetry. The number of unique entries is therefore given by $n_8= n_{\mathrm{pair}}(n_{\mathrm{pair}}+1)/2$, where $n_{\mathrm{pair}}= N(N+1)/2$. The two-body snapshot matrix is therefore defined as,
\begin{align}
{B}_{\theta,[pqrs]}
\Leftrightarrow
g_{pqrs}({\theta}),
\qquad
{B}\in\mathbb{R}^{M\times n_8},
\label{eq:snapshot2}
\end{align}
where \([pqrs]\) labels one unique index. All inner products between two-body snapshots are computed using only these unique entries, so symmetry-related entries are not counted multiple times.

\subsection{Geometry SVD in a joint basis}
\label{sec:geomsvd}

At this stage, we have separate snapshot matrices for the one- and two-body integral tensors. While we could perform a singular value decomposition of each matrix independently, this would produce one set of basis vectors for the one-body terms and another set for the two-body terms. Although this approach can compress each tensor separately, it would require the circuit to prepare two different geometry-dependent decompositions. It would also prevent us from using a single set of basis vectors and BLISS coefficients when constructing the Hamiltonian block encoding. We find that this results in a less efficient block encoding of the ensemble. We therefore propose using the single joint basis obtained by defining the normalized Gram matrix,
\begin{align}
\mathcal{G}
=
\cos^2(w)
\frac{{A}{A}^{\!\top}}
{\operatorname{tr}\!\left({A}{A}^{\!\top}\right)}
+
\sin^2(w)
\frac{{B}{B}^{\!\top}}
{\operatorname{tr}\!\left({B}{B}^{\!\top}\right)}.
\label{eq:jointgram}
\end{align}
The angle $w$ sets the relative weight of the two channels, interpolating between a basis fixed entirely by the one-body channel ($w=0$), and one fixed entirely by the two-body ($w=\pi/2$) channel and could be optimized per ensemble. We adopt equal weighting, $w=\pi/4$, throughout; all numerical results reported below use this choice. Since each term has unit trace, both the one- and two-body channels contribute equally to the overall basis. This prevents the channel with the larger overall Frobenius norm from determining the basis by itself. The Gram matrix has dimension $M\times M$, which is much smaller than the covariance matrices ${A}^{\!\top}{A}$ and ${B}^{\!\top}{B}$ when the number of grid points is smaller than the number of tensor entries. Since $M\ll n_8$ for the systems considered here, we use this Gram-matrix formulation, which is also known as the method of snapshots~\cite{Sirovich1987Turbulence, Wang2016Proper}. The compact basis representation is obtained by diagonalizing $\mathcal{G}$,
\begin{align}
\mathcal{G} = \sum_{\ell=0} \omega_\ell\, V_\ell(\theta) V_\ell(\theta'),
\label{eq:jointgram_eig}
\end{align}
where $V_\ell(\theta)$ is an orthonormal eigenvector defined over the geometry grid and $\omega_\ell$ measures the importance of each joint geometry profile, ordered so that $\omega_0\ge\omega_1\ge\cdots$. The one- and two-body mode tensors are obtained by projecting the snapshot matrices onto the joint basis,
\begin{align}
\mathcal{K}^{(\ell)}_{pq} &= \sum_\theta V_\ell(\theta)\,{A}_{\theta,pq} , \qquad G^{(\ell)}_{pqrs} = \sum_\theta V_\ell(\theta)\,{B}_{\theta,pqrs} .
\label{eq:modeproj}
\end{align}
The corresponding reconstruction using the leading $K$ basis modes is then given by,
\begin{align}
\kappa_{pq}({\theta}) &\approx \sum_{\ell=0}^{K-1}  V_\ell(\theta) \, \mathcal{K}^{(\ell)}_{pq}, \qquad g_{pqrs}({\theta}) \approx \sum_{\ell=0}^{K-1} V_\ell(\theta) \,G^{(\ell)}_{pqrs}.
\label{eq:modeexp}
\end{align}
These relations define the proposed compressed basis representation.

\subsection{Effective mode Hamiltonians}
\label{sec:modeham}

By substituting Eq.~\eqref{eq:modeexp} into the parameterized Majorana Hamiltonian $H({\theta})$, we obtain a geometry-independent chemistry-like Hamiltonian for each pair of mode tensors $\big(\mathcal{K}^{(\ell)},G^{(\ell)}\big)$:
\begin{align}
\widetilde{H}_\ell
=
\frac{i}{2}
\sum_{pq,\sigma}
\mathcal{K}^{(\ell)}_{pq}
\hat\gamma_{p\sigma,0}
\hat\gamma_{q\sigma,1}
-
\frac{1}{8}
\sum_{pqrs,\sigma\tau}
G^{(\ell)}_{pqrs}
\hat\gamma_{p\sigma,0}
\hat\gamma_{q\sigma,1}
\hat\gamma_{r\tau,0}
\hat\gamma_{s\tau,1}.
\label{eq:mode_operator}
\end{align}
The full parameterized Hamiltonian is therefore,
\begin{align}
H_{\mathrm{SVD}}({\theta})
=
\mathcal{E}_{{\theta}}
+
\sum_{\ell=0}^{K-1}
V_\ell({\theta})\,
\widetilde{H}_\ell,
\label{eq:modeham}
\end{align}
where all geometry dependence in the operator component is carried by the mode vectors $V_\ell({\theta})$. The scalar $\mathcal{E}_{{\theta}}$ is the additive nuclear repulsion and frozen-core energy contribution defined in Sec.~\ref{sec:param}. $\mathcal{E}_\theta$ remains outside the block encoding and is restored through the phase register when total energies are required. Sec.~\ref{sec:threshold} describes this procedure, and shows that the constant may be used to replace the approximate Hartree-Fock energy of the compressed Hamiltonian with the exact one. Since each $\widetilde{H}_\ell$ has the structure of a chemistry Hamiltonian, we may apply tensor factorization techniques to each mode independently. Here, we use BLISS-THC to compress the four-index tensors $G^{(\ell)}_{pqrs}$ into mode-dependent core tensors $\tilde{\zeta}^{(\ell)}_{\mu\nu}$ and leaf tensors $\chi^{\ell\mu}$. The one-body terms are represented by the corresponding shifted eigenvalues $\tilde{t}^{(\ell)}_k$. Here, the tilde denotes the unnormalized quantities returned by the BLISS-THC fit. Evaluating Eq.~\eqref{eq:lam_thc} for each mode gives the corresponding block encoding 1-norm,
\begin{align}
\sigma_\ell
=
\sum_k
\left|\tilde{t}^{(\ell)}_k\right|
+
\frac{1}{2}
\sum_{\mu,\nu}
\left|\tilde{\zeta}^{(\ell)}_{\mu\nu}\right|
-
\frac{1}{4}
\sum_{\mu}
\left|\tilde{\zeta}^{(\ell)}_{\mu\mu}\right|.
\label{eq:sigma}
\end{align}
Writing $\widetilde{H}^{\mathrm{BI}}_\ell$ for the operator that these fitted quantities reconstruct, that is, Eq.~\eqref{eq:mode_operator} with $\mathcal{K}^{(\ell)}$ and $G^{(\ell)}$ replaced by their block-invariant BLISS-THC representations, we define the normalized mode Hamiltonian $H_\ell = \widetilde{H}^{\mathrm{BI}}_\ell/\sigma_\ell$, which has unit 1-norm by construction. We note that $\tilde{H}_\ell$ is the exact mode Hamiltonian of Eq.~\eqref{eq:modeham} obtained from the SVD, whereas $\tilde{H}^\mathrm{BI}_\ell$ is its BLISS-THC approximation. Since the BLISS symmetry shift vanishes in the fixed $\eta$-electron sector, Eq.~\eqref{eq:modeham} becomes
\begin{align}
H_{\mathrm{SVD}}({\theta})
=
\mathcal{E}_{{\theta}}
+
\sum_{\ell=0}^{K-1}
\sigma_\ell
V_\ell({\theta})
H_\ell.
\label{eq:svd_ham}
\end{align}
This is the representation that we block encode below. The operators $H_\ell$ contain the compressed operator structure, $\sigma_\ell$ sets the energy scale of each mode, and $V_\ell({\theta})$ contains its geometry dependence.

\section{Quantum circuit}
\label{sec:circuits}

To encode the electronic structure of various geometries in the same wave function, we introduce separate registers: a physical register $\mathsf{s}$, representing an electronic system on $2N$ qubits, and an angle register $\mathsf{a}$ that stores the flattened geometry index $\theta$ as computational state $|\theta\rangle$. A typical state could look like 

\begin{align}
    |\Phi\rangle = \sum_{\theta} c_{\theta} |\theta\rangle_{\mathsf{a}}|\phi_{\theta}\rangle_{\mathsf{s}}\, , 
    \label{eq:typical_state}
\end{align}
where $c_\theta$ are coefficients and $|\phi_\theta\rangle$ are some states representing an electronic wave function. 

Following the arguments of Sec.~\ref{sec:svdthc}, we will address the subspace of every geometry $\theta$ encoded in the ensemble with the parameterized Hamiltonian $\hat{H}_\theta = \sum_\ell \sigma_\ell V_\ell(\theta) H_\ell$. For electronic structure calculations over ensembles of geometries, we require a Hamiltonian \emph{multiplexor}, a quantum circuit that assigns every $|\theta\rangle$ subspace of $\mathsf{a}$ its own Hamiltonian $\hat{H}_\theta$:

\begin{align}
\mux_\theta \hat{H}_\theta = 
\sum_{\theta}|\theta\rangle\!\langle \theta|_{\mathsf{a}}\otimes \hat{H}_{\theta}\, .
\label{eq:ensemble_ham}
\end{align}
Throughout this paper, we will use the notation $\mux_i O_i=\sum_i|i\rangle\!\langle i|\otimes O_i$ for various elements $O_i$ and indices $i$ in some auxiliary registers.

In this section, we define the SVD-THC circuit, a quantum program for $\mux_\theta \mathcal{B}[\hat{H}_\theta/\lambda ]$, the multiplexed block encoding $\mathcal{B}[\hat{H}_\theta/\lambda]$ of the Hamiltonians $\hat{H}_\theta$ normalized by a universal 1-norm $\lambda$. Using the special structure of $\hat{H}_\theta$, the SVD-THC circuit only multiplexes over $\theta$ values for the preparation of an auxiliary state $\propto\sum_\ell \sqrt{\sigma_\ell |V_\ell|}|\ell\rangle$. In an implementation for $\mux_\ell \mathcal{B}[H_\ell]$, we compress the block encodings $\mathcal{B}[H_\ell]$ of the SVD Hamiltonians $H_\ell$ with BLISS-THC. The objective is for the SVD-THC circuit to be cheaper than a circuit $\mux_{\theta} \mathcal{B}[H^\prime_\theta/\lambda_{\mathrm{max}}]$, where the Hamiltonians $H^\prime_\theta= H(\theta) - \mathcal{E}_\theta$ based on the uncompressed system in Eq.~\eqref{eq:ham} are block-encoded with BLISS-THC using its own universal 1-norm $\lambda_{\mathrm{max}}$, and where the multiplexor over $\theta$ is implemented through \emph{slicing}, i.e., through a sequence of gates

\begin{align}
\label{eq:slicing}
|\theta\rangle\!\langle\theta|_{\mathsf{a}}\otimes \mathcal{B}[H^\prime_\theta/\lambda_{\mathrm{max}}  ] + \sum_{\theta^\prime\!\neq \theta} |\theta^\prime\rangle\!\langle\theta^\prime|_{\mathsf{a}}\otimes 1 \, 
\end{align}

for $\theta = 0\dots(M-1)$. This \emph{sliced-THC} algorithm will serve as a baseline for the compression achieved using SVD-THC. Next, we will develop the circuit of SVD-THC, focusing first on the broad strokes including the preparation of the auxiliary state, before turning our attention to the implementation of $\mux_\ell \mathcal{B}[H_\ell]$.

\subsection{SVD-THC circuit}
\begin{table}[b]
    \centering
\begin{tabular}{rcl}
\hline \rule{0pt}{10pt} 
    $\mathsf{a}$ &...& \begin{minipage}{.8\textwidth}\flushleft Angle register of $O(\log M)$ qubits, holding the index $|\theta\rangle$. \end{minipage}   \\[2pt]
     $\mathsf{b}$ &...& \begin{minipage}{.8\textwidth}\flushleft SVD register, holding the index $|\ell\rangle$ in $O(\log K)$ qubits.   \end{minipage}\\[2pt]
$\mathsf{f}$ &...& \begin{minipage}{.8\textwidth}\flushleft Register holding a phase gradient state for catalytic rotation synthesis.  \end{minipage}\\[2pt]
     $\mathsf{g}$&...&\begin{minipage}{.8\textwidth}\flushleft Auxiliary register for the outer state preparation $\mathsf{prep}_{\mathsf{o},\theta}$. \end{minipage}\\[2pt]
     $\mathsf{j}$&...& \begin{minipage}{.8\textwidth}\flushleft Auxiliary register for the inner state preparation  $\mathsf{prep}_{\mathsf{i},\ell}$.\end{minipage} \\[2pt]
     $\mathsf{m}$&...& \begin{minipage}{.8\textwidth}\flushleft Temporary qubit encoding the signs of $V_\ell(\theta)$. \end{minipage}\\[2pt]
     $\mathsf{q}$&...& \begin{minipage}{.8\textwidth}\flushleft Auxiliary register for the THC circuit. \end{minipage} \\[2pt]
        $\mathsf{q}_1$ &...& \begin{minipage}{.8\textwidth}\flushleft Temporary qubit switching between 1- and 2-body operators in the THC circuit.\end{minipage} \\[2pt]
      $\mathsf{q}_4$, $\mathsf{q}_5$&...& \begin{minipage}{.8\textwidth}\flushleft Auxiliary registers of $\log(\max_\ell R_\ell)$ qubits each, holding the THC indices $|\mu\rangle$, $|\nu\rangle$ respectively.\end{minipage} \\[8pt]
     $\mathsf{s}$&...& \begin{minipage}{.8\textwidth}\flushleft $2N$-qubit register representing spin-orbitals of the molecular system. \end{minipage}  \\[2pt]
    $\mathsf{u}$&...& \begin{minipage}{.8\textwidth}\flushleft Auxiliary qubit in the Generalized Quantum Signal Processing~\cite{motlagh2024generalized} routine. \end{minipage} \\[2pt]
    $\mathsf{v}$ &...& \begin{minipage}{.8\textwidth}\flushleft Flag qubit padding the block encoding $\mathcal{B}[\hat{H}_\theta/\lambda]$ to the 1-norm $\lambda$.\end{minipage} \\[2pt]
    $\mathsf{w}$ &...& \begin{minipage}{.8\textwidth}\flushleft Flag qubit for the thresholding circuit of Fig.~$\ref{fig:thresh}$. \end{minipage}\\[2pt]
    $\mathsf{x}$ &...& \begin{minipage}{.8\textwidth}\flushleft Union of all permanent qubits of $\mathsf{b}$, $\mathsf{g}$, $\mathsf{j}$, $\mathsf{q}$ as well as $\mathsf{v}$.\end{minipage}\\[2pt]
    \hline
\end{tabular}
    \caption{Comprehensive list of all quantum registers encountered in this work.}
    \label{tab:registers}
\end{table}

\begin{figure}
    \centering
    \includegraphics[width=0.9\linewidth]{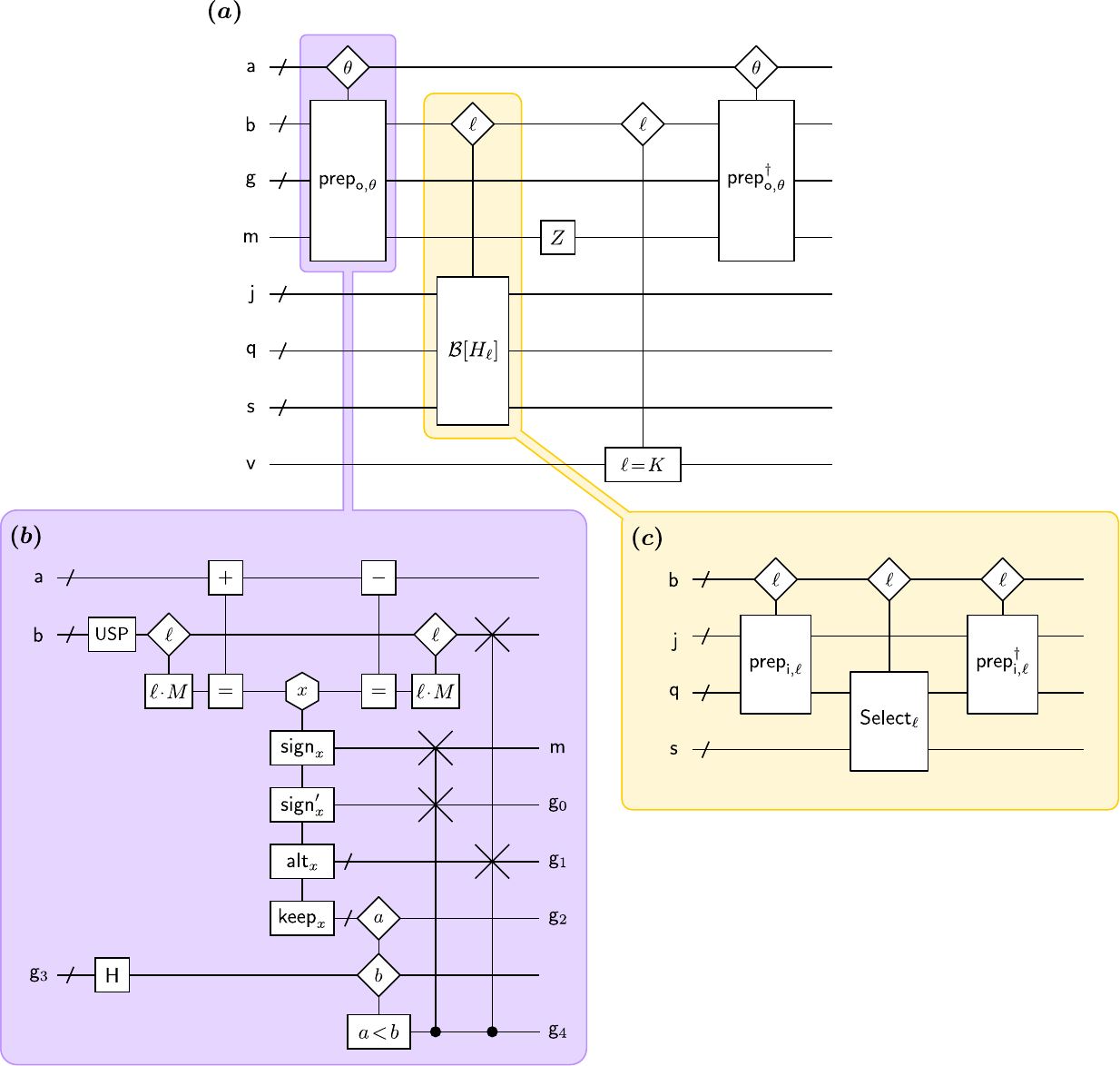}
    \caption{$\boldsymbol{(a)}$ SVD-THC circuit for $\mux_\theta \mathcal{B}[\hat{H}_\theta/\lambda]$, with respect to angle indices $\theta$ in register $\mathsf{a}$. The subroutine labeled `$\ell = K$' flips a qubit $\mathsf{v}$ in the $|K\rangle$ subspace of register $\mathsf{b}$. Diamond-shaped controls are our notation for general gate inputs. The $Z$ gate on $\mathsf{m}$ encodes a phase. All quantum registers are listed in Table~\ref{tab:registers}. $\boldsymbol{(b)}$ Outer state preparation. A $\theta$-multiplexed version of Alias Sampling~\cite{babbush2018encoding}. After a uniform state preparation $\mathsf{USP}$ on the $\mathsf{b}$ register, a joint index $x$ is computed in order 
    to load index aliases $\mathsf{alt}_x$ and coefficient data $\mathsf{keep}_x$, along with signs $\mathsf{sign}_x$ and their aliases $\mathsf{sign}^\prime_x$ efficiently. To this end, within the $|\ell\rangle$ subspace of $\mathsf{b}$, a temporary register is initialized in $|\ell M\rangle$, and the $\theta$ values in $\mathsf{a}$ are added to it. In-place adders (subtractors) add (subtract) the values of the registers under the gates labeled `$+$' (`$-$') to (from) the values in the registers under the gates labeled `$=$'. Hexagon-shaped gates denote data-loader inputs. The gate labeled `$a<b$' is a comparator, flipping a bit depending on the values in $\mathsf{g}_2$ and $\mathsf{g}_3$. `$\mathsf{H}$' denotes a tower of Hadamard gates on all qubits in register $\mathsf{g}_3$. From this circuit we learn that all sub-registers of $\mathsf{g}$ except $\mathsf{g}_3$, as well as qubit $\mathsf{m}$ can be considered temporary. $\boldsymbol{(c)}$ THC circuits multiplexed over the $\ell$ index in $\mathsf{b}$, featuring a modified Alias Sampling routine $\mathsf{prep}_{\mathsf{i},\ell}$ that uses a junk register $\mathsf{j}$ to prepare a block encoding register $\mathsf{q}$ for $\mathsf{Select}_\ell$.}
    \label{fig:blockencs}
\end{figure}

The SVD-THC circuit, our compressed implementation of the ensemble block encoding $\mux_\theta \mathcal{B}[\hat{H}_\theta/\lambda]$, is depicted in Figure~\ref{fig:blockencs}$(a)$.  Within this algorithm, a state preparation routine $\mathsf{prep}_{\mathsf{o},\theta}$ encodes features of the Hamiltonians $\hat{H}_\theta$ in a $\log(K+1)$-qubit register $\mathsf{b}$, a single qubit $\mathsf{m}$, as well as a garbage register $\mathsf{g}$. A comprehensive list of all quantum registers in this paper can be found in Table~\ref{tab:registers}. In the $|\theta\rangle$ subspace of register $\mathsf{a}$, $\mathsf{prep}_{\mathsf{o},\theta}$ prepares the state 

\begin{align}
\label{eq:oprep}
\mathsf{prep}_{\mathsf{o},\theta}|0\rangle_{\mathsf{bgm}} =  \sum_{\ell=0}^{K} \sqrt{\left|\xi_{\theta,\ell}\right|} \; |\ell\rangle_{\mathsf{b}}|\mathcal{Y}_{\theta,\ell}\rangle_{\mathsf{g}}|\mathsf{sign}_{\theta,\ell}\rangle_{\mathsf{m}}\, ,
\end{align}

in which $|\ell\rangle$ are computational basis states, $|\mathcal{Y}_{\theta,\ell}\rangle$ are irrelevant garbage states and bits $\mathsf{sign_{\theta,\ell}} \in \lbrace 0,1\rbrace$ encode signs of some coefficients $\xi_{\theta,\ell}$, such that 
\begin{align}
    (-1)^{\mathsf{sign}_{\theta,\ell}} |\xi_{\theta,\ell}| = \xi_{\theta,\ell}\, .
\end{align}
The coefficients themselves are chosen to match the SVD parameters of the Hamiltonian in Eq.~\eqref{eq:svd_ham} via 
\begin{align}
\xi_{\theta,\ell} = \sigma_\ell V_\ell(\theta) /\lambda\, ,
\end{align}
for $\ell \in [0, K-1]$. Since a Hamiltonian $\sum_\ell \xi_{\theta,\ell} H_\ell$  may have a different 1-norm at every value of $\theta$ due to the $\theta$-dependent outcome of the sum $\sum_\ell \sigma_\ell |V_\ell(\theta)|$, we use the subspace $|K\rangle$ of register $\mathsf{b}$ as padding. With what is essentially a large Toffoli gate in the center of Figure~\ref{fig:blockencs}$(a)$,  we flip a qubit $\mathsf{v}$ in the subspace of $\ell=K$, thus removing the amplitude $\xi_{\theta,K}$ from the operator encoded in the $|0\rangle$ subspace of  $\mathsf{m}$, $\mathsf{v}$, $\mathsf{g}$, as well as the auxiliary registers $\mathsf{j}$ and $\mathsf{q}$. The Hamiltonian
\begin{align}
 \langle 0 | \mathcal{B}[\hat{H}_\theta/\lambda]|0\rangle_{\mathsf{gjmqv}} = \sum_{\ell=0}^{K-1} \xi_{\theta, \ell} H_\ell \, 
\end{align}
is therefore subnormalized by the $\theta$-dependent factor, $1-|\xi_{\theta,K}|$. There is some value $\theta^*$ of $\theta$ that realizes the maximum 1-norm for which no padding is needed such that the $|K\rangle$ subspace of $\mathsf{b}$ vanishes: $\xi_{\theta^*,K}=0$. The 1-norm of the SVD-THC is thus, 
 \begin{align}
\label{eq:svdthc1norm}
    \lambda = \max_{\theta} \sum_{\ell=0}^{K-1} \sigma_\ell \left|V_\ell(\theta) \right| = \sum_{\ell=0}^{K-1}\sigma_\ell\left|V_\ell(\theta^*)\right|\, .
\end{align}
The circuit for $\mux_\theta\mathsf{prep}_{\mathsf{o},\theta}$, depicted in Figure~\ref{fig:blockencs}$(b)$, is a multiplexed version of quantum Alias Sampling. In the $|\theta\rangle$-subspace of register $\mathsf{a}$, Alias Sampling subjects the uniform state $\propto\sum_\ell |\ell\rangle$ over all the indices $\ell=0, ..., K$, to a subroutine that retains only the $(\mathsf{keep}_{\theta,\ell})$-portion of each index $|\ell\rangle$ and assigns the rest to an alias index $\mathsf{alt}_{\theta,\ell}$. Here $\mathsf{keep}_{\theta,\ell}\in[0,1)$  represent fixed-point numbers, while $\mathsf{alt}_{\theta,\ell}\in [0,K]$ are integers. In the process, it also sets the sign qubit $\mathsf{m}$ with the values $\mathsf{sign}_{\theta,\ell}$  and  $\mathsf{sign}_{\theta,\ell}^\prime$ in the kept and re-assigned subspaces, respectively. Each index $|\ell\rangle$ turns into

\begin{align}
|\ell\rangle_{\mathsf{b}}|0\rangle_{\mathsf{g}}|0\rangle_{\mathsf{m}} \mapsto \sqrt{\mathsf{keep}_{\theta,\ell}}  |\ell\rangle_{\mathsf{b}} |\mathcal{Z}_{\theta,\ell,0}\rangle_{\mathsf{g}}|\mathsf{sign}_{\theta,\ell}\rangle_{\mathsf{m}}+ \sqrt{1-\mathsf{keep}_{\theta,\ell}}|\mathsf{alt}_{\theta,\ell}\rangle_{\mathsf{b}}|\mathcal{Z}_{\theta,\ell, 1}\rangle_{\mathsf{g}}|\mathsf{sign}_{\theta,\ell}^\prime\rangle_{\mathsf{m}}\, ,
\end{align}

where  $|\mathcal{Z}_{\theta,\ell,i}\rangle$ are garbage states.  
The values of $\mathsf{alt}$, $\mathsf{keep}$, $\mathsf{sign}$ and $\mathsf{sign}^\prime$ are chosen appropriately to encode the coefficients $\xi_{\theta,\ell}$ and signs in Eq.~\eqref{eq:oprep}. Within the $\mux_\theta \mathsf{prep}_{\mathsf{o},\theta}$ circuit, all these values are loaded into registers $\mathsf{g}$ and $\mathsf{m}$ from the index pairs in registers $\mathsf{a}$ and $\mathsf{b}$.
This double-indexing can cause a cost overhead in data-loading, which forces us to find a joint index $x(\ell, \theta)$. Since we are expecting neither $K+1$ nor $M$ to be an integer power of two, i.e., a number $2^k$ with integer $k$, the data-loader in the circuit of Figure~\ref{fig:blockencs}$(b)$ would have an $O(K \log M)$ overhead of Toffoli gates (left elbows).

QROM circuits have a Toffoli cost linear in the number of payload items only when they load from a continuous index~\cite{childs2018toward,babbush2018encoding}. When loading from $\mathsf{a}$ and $\mathsf{b}$ as if they were one giant register, the QROM would have to iterate over binary numbers $\theta \oplus \ell$, formed by appending indices $\theta$ and $\ell$, which are not continuous. 
Since the integers $(M-1) \oplus \ell$ and $0\oplus (\ell+1)$ do not differ by 1, every interface between pairs of adjacent indices $(M-1)\oplus \ell$ and $0 \oplus (\ell+1)$ incurs  $O(\log M)$ Toffoli gates in the unitary iteration procedure. 

With the data-loader already being the dominant cost of  $\mux_\theta \mathsf{prep}_{\mathsf{o},\theta}$, we want to avoid this scenario by joining $\ell$ and $\theta$ into a continuous index $x(\ell, \theta) = M\ell + \theta$. To this end we use a trick from Ref.~\cite{von2021quantum}, and first load the \emph{basis} of the more significant index in $x$ -- meaning here the $\ell$-dependent part of $x$  -- into a temporary register. With the basis  $M\ell$ being linear in $\ell$, this temporary register may be initialized in $|M\ell\rangle$ by a multiplication circuit rather than a data-loader. In the second step, the joint index $|x(\ell, \theta)\rangle$ is finalized by adding the value of the less significant index, here $\theta$,  into the temporary register, simply done with an in-place adder originating from register $\mathsf{a}$. Basis-loading and addition have complexity $O(K+\log M)$, which is preferable to the $O(K\log M)$ overhead incurred by loading $\mathsf{alt}$, $\mathsf{keep}$, $\mathsf{sign}$ and $\mathsf{sign}^\prime$ values from a discontinuous index.

\subsection{Multiplexed THC circuit}
The SVD-THC circuit still contains a block encoding multiplexor $\mux_\ell \mathcal{B}[H_\ell]$, and while we expect this subroutine, depicted in Figure~\ref{fig:blockencs}$(c)$, to be the most expensive part of the entire circuit, its complexity should be greatly reduced compared to the sliced-THC algorithm.

For one, we are multiplexing THC Hamiltonians over the index $\ell$  which has a smaller range than $\theta$. One can hope for a compression when $K<M$ assuming comparable THC ranks and parameters. We also do not want to multiplex $\mathcal{B}[H_\ell]$ by slicing. Instead, we modify the THC circuit of Ref.~\cite{caesura2025faster}, such that we share subroutines such as the Givens rotations between all block encodings $\mathcal{B}[H_\ell]$, and simply extend internal data-loaders by another index $\ell$. 

One challenge here is to account for Hamiltonians $H_\ell$ having individual THC ranks. The THC rank $R_\ell$ of $H_\ell$ is a compression parameter of the block encoding. In fact,  a leading factor in the complexity of the THC circuit is $O(\mathrm{poly}(R_\ell))$. In this paper, we construct a multiplexed THC circuit, such that Hamiltonians with smaller rank contribute less to the overall complexity, striving for a scaling of $\sum_\ell \mathrm{poly}(R_\ell) $ rather than $K \times \mathrm{poly}\left(\max_\ell R_\ell\right)$.

The THC circuits encode the Hamiltonians

\begin{align}
\label{eq:thcham}
    H_\ell\; = \;    \frac{1}{8}\sum_{\mu=0}^{R_\ell-1} & \sum_{\substack{\nu=0\\ \nu  \neq \mu}}^{R_\ell-1} \zeta_{\ell\mu\nu} \sum_{\sigma,\tau = \uparrow, \downarrow}\,Z_\sigma\!\left[\boldsymbol{\hat{\vartheta}}_{\ell\mu} \right]\; Z_\tau\!\left[\boldsymbol{\hat{\vartheta}}_{\ell\nu}\right] \\ + \frac{1}{2} &\sum_{\mu=0}^{R_\ell-1} \zeta_{\ell\mu\mu} \sum_{\substack{\sigma,\tau=\uparrow,\downarrow\\ \sigma \neq \tau}} \;Z_\sigma\!\left[\boldsymbol{\hat{\vartheta}}_{\ell\mu}\right]\; Z_{\tau}\!\left[\boldsymbol{\hat{\vartheta}}_{\ell\mu}\right]\\ + \frac{1}{2}&\sum_{\mu=0}^{N-1}\zeta_{\ell\mu R_\ell} \sum_{\sigma=\uparrow,\downarrow}\,Z_\sigma\!\left[\boldsymbol{\widetilde{\vartheta}}_{\ell\mu} \right] 
\end{align}
with normalized `machine' coefficients  $\zeta_{\ell\mu\nu}$ such that $\sum_{\mu\leq \nu}|\zeta_{\ell\mu\nu}|=1$, and where $\zeta_{\ell\mu\nu} = \zeta_{\ell\nu\mu}$ for $\mu,\nu<R_\ell$. The  Hamiltonian features rotated occupation operators
\begin{align}
Z_\sigma[\boldsymbol{\vartheta}] = G_\sigma^\dagger(\boldsymbol{\vartheta}) \left(1-2a^\dagger_{0,\sigma}a^{\phantom{\dagger}}_{0,\sigma}\right)G_{\sigma}(\boldsymbol{\vartheta})
\end{align}
parameterized by sets of angles $\boldsymbol{\vartheta} = \vartheta_0,\vartheta_{1},\cdots,\vartheta_{N-2}$,  and Givens rotations 
\begin{align}
G_{\sigma}(\boldsymbol{\vartheta})=G_{\sigma,0}(\vartheta_0) \cdot G_{\sigma,1} (\vartheta_1) \cdots G_{\sigma,N-2}(\vartheta_{N-2})
\end{align}
with 
\begin{align}
    G_{\sigma,k}(\vartheta_k) = \exp\!\left[\frac{\vartheta_k}{2} \left(a^\dagger_{k+1,\sigma}a^{\phantom{\dagger}}_{k, \sigma} - a^\dagger_{k,\sigma}a^{\phantom{\dagger}}_{k+1,\sigma}\right)\right]\, .
\end{align}
The machine coefficients $\zeta_{\ell\mu\nu}$ and angles $\boldsymbol{\hat{\vartheta}}_{\ell\mu}$, $\boldsymbol{\widetilde{\vartheta}}_{\ell\mu}$ are matched with the parameters $\tilde{t}^{(\ell)}_k$ and $\tilde{\zeta}_{\mu\nu}^{(\ell)}$ and $\chi^{\ell\mu}$ from Section~\ref{sec:svdthc} as

\begin{align}
    \zeta_{\ell\mu\nu} = \frac{1}{\sigma_\ell}\begin{cases}
        -\tilde{t}_\mu^{(\ell)} & \text{for}\;  \nu=R_\ell,\, \mu<N; \\
        \tilde{\zeta}^{(\ell)}_{\mu\nu} & \text{for }\; \nu\in [0, R_\ell),\, \mu\in[0,\nu]; \\
        \tilde{\zeta}^{(\ell)}_{\mu\mu}/4& \text{for}\; \nu \in [0,R_\ell), \, \mu=\nu; \\
        0 & \text{else,}
    \end{cases}
\end{align}
 and 
 \begin{align}
     \chi^{\ell\mu}_{k}= \cos(2\hat{\vartheta}_{\ell \mu k}) \prod_{m<k} \sin(2\hat{\vartheta}_{\ell \mu m})
 \end{align}
for $k\in [0,N-2]$ as well as $\chi^{\ell\mu}_{N-1} = \prod_k\sin(2\hat{\vartheta}_{\ell \mu k})$, where $\hat{\vartheta}_{\ell\mu k}$ are the components of $\boldsymbol{ \hat{\vartheta}}_{\ell\mu}$.

The parameters $\boldsymbol{\vartheta}$ split into two groups: a group of angles $\boldsymbol{\hat{\vartheta}}_{\ell\mu}$ that encode two-body operators, and the groups of angles $\boldsymbol{\widetilde{\vartheta}}_{\ell\mu}$ encoding one-body operators. In addition to the superindex $\ell$, two-body angles depend on  THC indices $\mu$ and $\nu$ in the range from $0$ to $R_\ell-1$, while one-body angles depend on orbital indices in the range of $[0,N-1]$. 

The block encoding circuit is decomposed into an inner $\mathsf{prep}$ and $\mathsf{Select}$ circuit:

\begin{align}
\mathcal{B}[H_\ell]=\mathsf{prep}_{\mathsf{i},\ell}^\dagger\;  \mathsf{Select}_\ell \;\mathsf{prep}_{\mathsf{i},\ell}^{\vphantom{\dagger}}\, .
\end{align}

A circuit for $\mathsf{prep}_{\mathsf{i},\ell}$ is presented in Appendix~\ref{sec:prepare}. While somewhat similar to the multiplexed Alias Sampling of $\mathsf{prep}_{\mathsf{o},\theta}$, the inner $\mathsf{prep}$ has some more intricate design choices such as a multiplexed $\mathsf{USP}$. In the $|\ell\rangle$ subspace of $\mathsf{b}$, $\mathsf{prep}_{\mathsf{i},\ell}$ prepares registers $\mathsf{j}$  and $\mathsf{q}$ in the state
\begin{align}
\label{eq:prepstate}
\mathsf{prep}_{\mathsf{i},\ell}|0\rangle_{\mathsf{jq}} = \sum_{\nu = 0}^{R_\ell}\sum_{\mu\leq\nu} \sqrt{\left|\zeta_{\ell\mu\nu}\right|} \, |\mu\rangle_{\mathsf{q}_4}|\nu\rangle_{\mathsf{q}_5} |\Omega_{\ell\mu\nu}\rangle
\end{align}
with the main indices $|\mu\rangle$ and $|\nu\rangle$ encoded in registers $\mathsf{q}_4$ and $\mathsf{q}_5$, respectively. The state $|\Omega_{\ell\mu\nu}\rangle$ encodes additional information: a success flag on qubit $\mathsf{q}_0$, a flag for the one-body terms in the case $\nu=R_\ell$ on qubit $\mathsf{q}_1$, the sign of $\zeta_{\ell\mu\nu}$ in qubit $\mathsf{q}_2$, as well as a junk state on register $\mathsf{j}$.

\begin{sidewaysfigure}
    \centering
    \includegraphics[width=.8\linewidth]{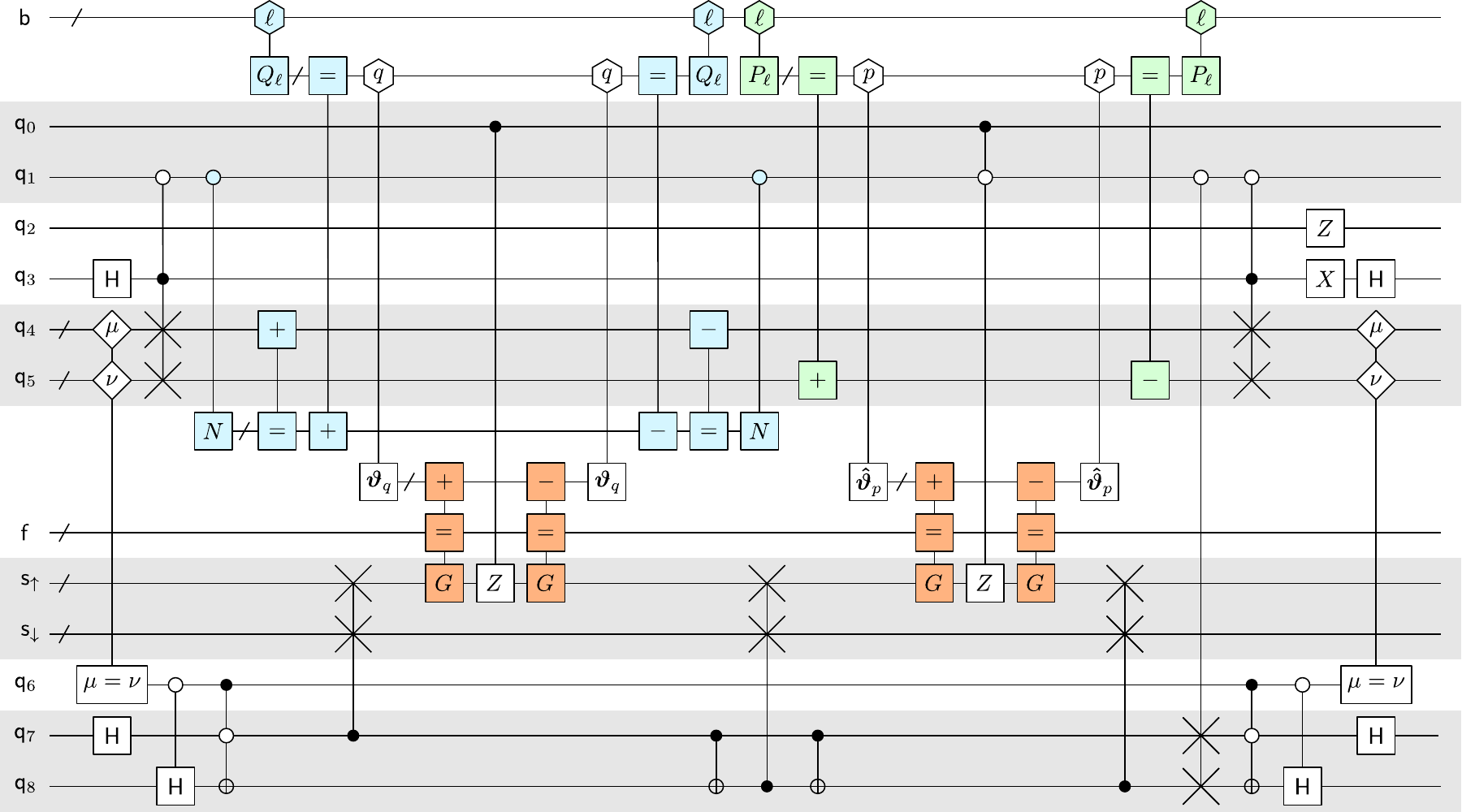}
    \caption{Quantum circuit for $\mux_\ell \mathsf{Select}_\ell$. This circuit loads a set of different rotation angles $\boldsymbol{\hat{\vartheta}}_p$ and  $\boldsymbol{\vartheta}_q$ that serve as the input for sequences of Givens rotations (orange) implementing the rotated operators $\mux_{p} Z_{\sigma}[\boldsymbol{\hat{\vartheta}}_p]$ and $\mux_q Z[\boldsymbol{\vartheta}_q]$ respectively, using in-place adders/subtractors into a phase gradient state in register $\mathsf{f}$. Here $p(\ell,\nu)$ and $q(\ell,b,\mu)$ are contiguized numbers from indices $(\ell, b,\mu,\nu)$ in registers $(\mathsf{b},\mathsf{q}_1,\mathsf{q}_4,\mathsf{q}_5)$. Since the value $\ell$ in $\mathsf{b}$ determines the range of $\mu$ and $\nu$, we load $\ell$-dependent bases $Q_\ell$ and $P_\ell$ into  respective temporary registers, which then receive values of $\mu + \bar{b}N$ (blue) and $\nu$ (green) via in-place addition, where $\bar{b}$ is the opposite value of $b$. Note that the value $b=1$ in $\mathsf{q}_1$ flags the subspace in which angles for one-body terms must be loaded. The remainder of the circuit follows the construction in \cite{caesura2025faster}, with qubit $\mathsf{q}_2$ holding information about the sign of the encoded operator, $\mathsf{q}_3$ is a sort of coin qubit making the circuit self-inverse, $\mathsf{q}_6$ flags the case where $\mu=\nu$ after a comparator while $\mathsf{q}_7$ and $\mathsf{q}_8$ switch the spin-up $\mathsf{s}_\uparrow$ and spin-down parts $\mathsf{s}_\downarrow$ of the simulator register.}
    \label{fig:mux_select}
\end{sidewaysfigure}

The multiplexed $\mathsf{Select}$ routine, depicted in Figure~\ref{fig:mux_select}, encodes two-body operators 
\begin{align}
Z_{\sigma}\!\left[\boldsymbol{\hat{\vartheta}}_{\ell\mu}\right]Z_{\tau}\!\left[\boldsymbol{\hat{\vartheta}}_{\ell\nu}\right] +  Z_{\sigma}\!\left[\boldsymbol{\hat{\vartheta}}_{\ell\nu}\right]Z_{\tau}\!\left[\boldsymbol{\hat{\vartheta}}_{\ell\mu}\right]
\end{align}
in the $|\mu\rangle|\nu\rangle$ subspace of $\mathsf{q_4}$ and $\mathsf{q_5}$ for $\mu,\nu \in [0, R_\ell)$. The circuit itself has two parts: the first half of the circuit encodes operator $\mux_{\ell \mu} Z_\sigma[\boldsymbol{\hat{\vartheta}}_{\ell \mu}]$ with  $\mu$ in $\mathsf{q}_4$ and $\ell$ in $\mathsf{b}$, while the second half implements  $\mux_{\ell \nu }Z_\sigma[\boldsymbol{\hat{\vartheta}}_{\ell\nu}]$ with $\nu$ in $\mathsf{q}_5$. Within the $|\ell\rangle|\nu\rangle$ subspace of their respective registers, the angles $\boldsymbol{\hat{\vartheta}}_{\ell \nu}$ are loaded into a temporary register $\mathsf{t}$, before the orange gates in Figure~\ref{fig:mux_select} encode the Givens rotations $G^{(\dagger)}_{\sigma}(\boldsymbol{\hat{\vartheta}}_{\ell\nu})$  by continuously adding different cells of $\mathsf{t}$ into a phase gradient state in $\mathsf{f}$~\cite{Kivlichan2018}. This procedure applies to both halves of the circuit, but the data-loading of the second half is more straightforward. Rather than loading data $\boldsymbol{\hat{\vartheta}}_{\ell\nu}$ from the double index $\ell \oplus \nu$, we will load the same angles $\boldsymbol{\hat{\vartheta}}_{p}$ from a continuous index $p$ of the form
\begin{align}
    p(\ell, \nu) = P_\ell + \nu\, ,
\end{align}
where the basis $P_\ell = \sum_{m<\ell} R_m$ is loaded from $\mathsf{b}$ into another temporary register and the index $\nu$ from $\mathsf{q}_5$ is added to it. The corresponding gates are highlighted green in Figure~\ref{fig:mux_select}. Note that $\nu$ could be out of bounds: although the joint index $p$ is designed for $\nu$ to not exceed the  $\ell$-dependent values of $R_\ell-1$, the state in \eqref{eq:prepstate} allows for the indices $\nu=R_\ell$. However, in this case the bit $\mathsf{q}_1$ will be set, and the $Z$ operator between the Givens rotations is not applied, making the angles of the Givens rotations irrelevant. In this $|1\rangle|\mu\rangle|R_\ell\rangle$ subspace of the qubits $(\mathsf{q}_1, \mathsf{q}_4,\mathsf{q}_5)$, the second half of the multiplexed $\mathsf{Select}$ circuit is switched off entirely, and its first part encodes the one-body operators $\mux_{\ell\mu} Z_\sigma[\boldsymbol{\widetilde{\vartheta}}_{\ell\mu}]$ for $\mu\in [0,N)$. For the first half of the circuit this means also that, in addition to loading angles $\boldsymbol{\hat{\vartheta}}_{\ell\mu}$ in the $|0\rangle$ subspace of $\mathsf{q}_1$, angles $\boldsymbol{\widetilde{\vartheta}}_{\ell\mu}$ need to be loaded in the $|1\rangle$ subspace of $\mathsf{q}_1$. We combine  $\boldsymbol{\hat{\vartheta}}$ and $\boldsymbol{\widetilde{\vartheta}}$ into a triple-indexed set of parameters $\boldsymbol{\vartheta}_{\ell b \mu}$, where $|b\rangle$ is the index of $\mathsf{q}_1$. The values themselves will be loaded as $\boldsymbol{\vartheta}_{q}$ from a joint index $q$ with
\begin{align}
    q(\ell,b,\mu) = Q_\ell + \bar{b} N + \mu\, ,
\end{align}
 where $\bar{b} = b+1 \;\mathrm{mod}\, 2$, and $Q_\ell = \ell N + \sum_{m<\ell} R_m$. The gates involved in the computation of the joint index $q$ are highlighted light blue in Figure~\ref{fig:mux_select}, where we XOR the number $N$ into a temporary register, and then perform an in-place addition from register $\mathsf{q}_4$. The resulting values are  added to a different temporary register, into which the values of the basis $Q_\ell$ have been loaded from $\mathsf{b}$. The SVD-THC compression and circuit are deployed to molecular systems in the next sections.

\section{Compression results}
\label{sec:results}

In this section we present numerical results for three benchmark molecular systems, highlighting both the efficacy of the compressed block encoding and the challenges associated with it. We first introduce the molecules, the procedure used to generate their geometries, and the resulting potential energy surfaces. We then define the accuracy metric used throughout, examining how the compression error converges with the number of retained SVD modes. Next we compare the effective 1-norm of each mode with the bare norm that the factorization penalizes.

\subsection{Molecular systems and geometries}
\label{sec:systems}

We evaluate the parameterized compression scheme on three molecules. The first is 1-bromo-2-chloroethane (BCE), scanned over the $\mathrm{Cl}$--C--C--Br torsion $\varphi$ on a 73-point grid. The second is the glycine dipeptide, scanned over the backbone dihedrals $(\varphi,\psi)$ on a $37\times72$ grid. The third is the alanine dipeptide, scanned over the two backbone dihedrals of a $72\times72$ grid. The BCE torsion and the $\varphi$ axis of glycine dipeptide are sampled at Chebyshev-Gauss-Lobatto points on $[0^\circ,180^\circ]$, which cluster near the endpoints, and all other axes uniformly in steps of $5^\circ$.

\begin{figure}[ht!]
    \centering
    \includegraphics[width=\linewidth]{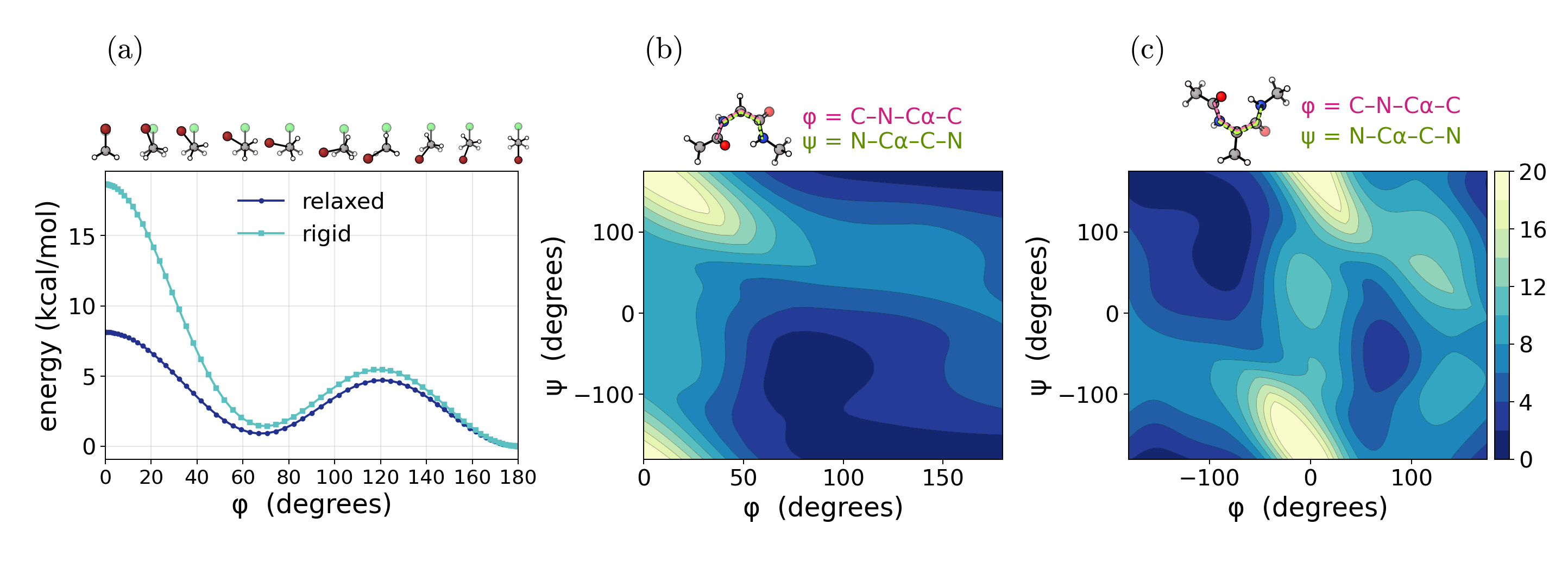}
    \caption{Potential energy surfaces (g-xTB) for (a) relaxed and rigid BCE, as well as the (b) glycine and (c) alanine dipeptides vs the backbone $(\varphi,\psi)$, on respectively $37\times72$ and $72\times72$ grids. Each panel carries an inset showing the molecule and the coordinate being scanned. Panel (a) shows a filmstrip of ten scanned geometries. Panels (b) and (c) show the two backbone dihedrals as colored dashed lines over the four atoms that define them; all remaining internal coordinates are drawn in gray and relaxed at every grid point. }
    \label{fig:surfaces}
\end{figure}

All molecular geometries were optimized using the g-xTB tight-binding method based on the xtb program~\cite{Froitzheim2025gxtb,Bannwarth2021xtb}, with harmonic restraints holding the scanned dihedrals and all other coordinates relaxed to very tight convergence. Fig.~\ref{fig:surfaces} shows the resulting potential energy surfaces. All correlated energies use the STO-3G basis. We freeze the core orbitals of a restricted Hartree-Fock (RHF) calculation and build the active space from the remaining valence atomic orbitals, projected against the core and L\"owdin orthogonalized~\cite{Lowdin1950}. This gives the frozen-core active spaces (26e, 20o) for BCE, (52e, 46o) for glycine dipeptide and (58e, 52o) for alanine dipeptide, in which correlation energies are computed with CCSD. Freezing the core affects relative energies along a torsion negligibly, since core orbitals are spatially compact and barely respond to the rotation. The minimal basis set remains the dominant approximation, and is simply used as a proof-of-concept.

Both BCE and the glycine dipeptide carry a mirror symmetry that allows us to halve the sampled domain. Rotating $\varphi$ to $-\varphi$ produces the mirror image of BCE, resulting in $E(\varphi)=E(-\varphi)$. The physical range therefore requires only $\varphi\in[0^\circ,180^\circ]$, which is the domain of Fig.~\ref{fig:surfaces}(a) and of the 73-point grid in Table~\ref{tab:ccsd}. Glycine dipeptide has no side chain, so its $\mathrm{C}_\alpha$ atom carries two equivalent hydrogens rather than a stereocenter. Its Ramachandran surface therefore possesses the analogous mirror symmetry $E(\varphi,\psi)=E(-\varphi,-\psi)$. Since $\psi$ is sampled over its full range, the point at $(-\varphi,\psi)$ coincides with the already sampled point at $(\varphi,-\psi)$, so only $\varphi$ must be halved. This gives the $37\times72$ grid of Table~\ref{tab:ccsd} rather than $72\times72$. The methyl side chain of alanine dipeptide makes its $\mathrm{C}_\alpha$ a stereocenter and breaks this symmetry, so its $72\times72$ grid samples the full surface in both angles.

For BCE, we compute both a relaxed scan, in which every other internal coordinate is re-optimized at each value of $\varphi$, and a rigid scan, in which only $\varphi$ is rotated while all other coordinates remain fixed at their global-minimum configuration values. Relaxation represents the physical scenario where the molecule can adjust its bond lengths, bond angles, and remaining torsions in response to the changing dihedral. A rigid scan prevents these adjustments and therefore overestimates the energy away from the minimum. Figure~\ref{fig:surfaces}(a) shows that this effect is substantial. The relaxed barrier is $8.1$~kcal/mol, compared with $18.7$~kcal/mol for the rigid scan. Evaluating CCSD in the active space on the same geometries gives barriers of $5.0$ and $18.7$~kcal/mol, respectively. The two levels of theory agree closely for the rigid scan but differ substantially once relaxation is allowed. 

\begin{table}[ht!]
\centering
\small
\setlength{\tabcolsep}{3.2pt}
\vspace{0.1cm}
\begin{tabular}{lrrrrrrrrr}
\toprule
 & & \multicolumn{4}{c}{g-xTB (mHa)} & \multicolumn{4}{c}{CCSD (mHa)} \\
\cmidrule(lr){3-6}\cmidrule(lr){7-10}
grid & 
$M$& min & mean & max & span & min & mean & max & span \\
\midrule
\multicolumn{10}{l}{\emph{1-bromo-2-chloroethane (BCE)}}\\
73, 1D $\varphi$ & 10 & 0.021 & 2.493 & 6.062 & 12.95 & 0.091 & 1.496 & 3.534 & 8.02 \\
 & 19 & 0.009 & 1.366 & 3.300 & 12.95 & 0.006 & 0.809 & 1.893 & 8.02 \\
 & 37 & 0.001 & 0.696 & 1.652 & 12.95 & 0.000 & 0.409 & 0.958 & 8.02 \\
 & 55 & 0.000 & 0.464 & 1.648 & 12.95 & 0.000 & 0.273 & 0.935 & 8.02 \\
 & 73 & 0.000 & 0.348 & 0.836 & 12.95 & 0.000 & 0.206 & 0.483 & 8.02 \\
\midrule
\multicolumn{10}{l}{\emph{Glycine dipeptide}}\\
10$\times$18 & 180 & 0.006 & 1.941 & 14.194 & 33.80 & 0.005 & 1.344 & 8.067 & 15.23 \\
19$\times$18 & 342 & 0.001 & 1.375 & 7.574 & 34.12 & 0.000 & 1.036 & 5.453 & 16.21 \\
19$\times$36 & 684 & 0.000 & 1.003 & 7.580 & 34.24 & 0.000 & 0.715 & 5.453 & 16.21 \\
37$\times$36 & 1,332 & 0.000 & 0.706 & 3.855 & 34.38 & 0.000 & 0.539 & 3.601 & 16.21 \\
37$\times$72 & 2,664 & 0.000 & 0.509 & 3.855 & 34.38 & 0.000 & 0.368 & 3.924 & 16.28 \\
\midrule
\multicolumn{10}{l}{\emph{Alanine dipeptide}}\\
3$\times$3 & 9 & 1.674 & 5.605 & 8.598 & 9.52 & 0.130 & 1.848 & 3.984 & 5.49 \\
9$\times$9 & 81 & 0.002 & 3.872 & 19.302 & 33.61 & 0.026 & 2.847 & 9.759 & 15.33 \\
18$\times$18 & 324 & 0.011 & 2.147 & 12.703 & 37.58 & 0.001 & 1.685 & 7.281 & 16.39 \\
36$\times$36 & 1,296 & 0.001 & 1.116 & 6.423 & 37.58 & 0.002 & 0.884 & 4.285 & 17.95 \\
72$\times$72 & 5,184 & 0.000 & 0.565 & 3.243 & 38.79 & 0.000 & 0.452 & 4.713 & 18.94 \\
\bottomrule
\end{tabular}
\caption{Energy landscape of the scanned geometries at two levels of theory. For each nested sub-grid, absolute energy differences between neighboring grid points are reported through the minimum, mean and maximum over all neighboring pairs, together with the energy span $E_{\max}-E_{\min}$. Total energies are evaluated on the same geometries with g-xTB, and CCSD with a frozen core.}
\label{tab:landscape}
\end{table}

\begin{figure}[t!]
    \centering
    \includegraphics[width=\linewidth]{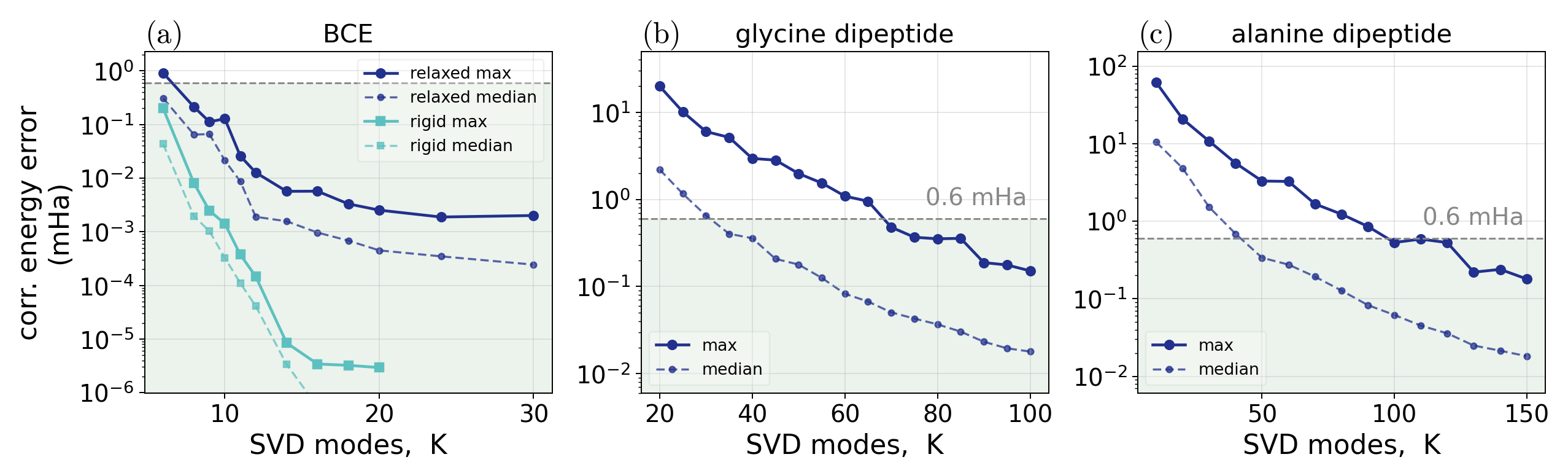}
    \caption{Correlation energy error versus retained SVD rank $K$ for (a) BCE including rigid and relaxed convergence, (b) relaxed glycine dipeptide, (c) relaxed alanine dipeptide. Gray dashed line is the $0.6$~mHa target. Dashed lines of the same color give the median error over the grid, added as a reference. Every point is evaluated at all geometries of the full grid.}
    \label{fig:convergence}
\end{figure}

The focus for this manuscript is primarily on relaxed potential energy surfaces because they provide an effective low-dimensional description of the energy landscape and they most closely follow the practical workflows in computational drug design. Once the remaining internal coordinates have been minimized at each value of $\varphi$, their energetic response is absorbed into the exposed coordinates rather than being represented explicitly. This reduction becomes especially valuable for higher-dimensional scans, where optimizing secondary coordinates can eliminate dimensions that would otherwise require additional Hamiltonian parameters. Relaxation can therefore simplify the overall representation, but it makes the associated integrals harder to compress, as illustrated by Figure~\ref{fig:convergence}(a). Reduced compressibility can also occur due to near-avoided crossings, competing minima, or changes in the identity of the optimized structure, where the minimizing geometry may vary rapidly or develop a cusp. For larger molecules such as the glycine and alanine dipeptides, rigid scans could introduce collisions between different parts of the molecule, and we therefore report relaxed surfaces only for the dipeptides, since they provide the more physically relevant energy landscapes.

Finally, since the proposed compression scheme is applied to a discretized surface, it is also useful to record how much the energy changes between adjacent grid points. Table~\ref{tab:landscape} reports this quantity for each full grid and its nested sub-grids, at both levels of theory. Apart from the $3\times3$ alanine dipeptide grid, which is too coarse to resolve the surface, the mean step between neighboring points decreases monotonically as the grid is refined, in every system and at both levels, and the maximum step follows the same trend less regularly. Refinement therefore resolves the existing surface rather than exposing new structure. Since variations below the error tolerance cannot be distinguished from the compression error, we treat the finest grids as resolving the relevant structure of the surfaces.

\FloatBarrier
\subsection{Accuracy metric}
\label{sec:accuracy}

Let $E^{\mathrm{CCSD}}_{\mathrm{corr}}(\cdot)$ denote the CCSD correlation energy of a Hamiltonian, and let $ H_{\mathrm{SVD}}$ denote compressed approximations obtained at rank $K$ as in Eq.~\eqref{eq:svd_ham} while $H(\boldsymbol{\theta_n})$ are the unfactorized Hamiltonians of Eq.~\eqref{eq:ham}. We measure the error as the worst case over the full
sampled grid,
\begin{align}
    \varepsilon_\infty(K) = \max_{\boldsymbol{n}} \left| E^{\mathrm{CCSD}}_{\mathrm{corr}}\!\big(H_{\mathrm{SVD}}(\boldsymbol{\theta_n})\big)
    - E^{\mathrm{CCSD}}_{\mathrm{corr}}\!\big(H(\boldsymbol{\theta_n})\big) \right|\, ,
    \label{eq:accuracy}
\end{align}
evaluated at every grid point. We compare correlation energies since the exact Hartree-Fock energy at each geometry can be restored by modifying the constant term $\mathcal{E}_\theta$ in Eq.~\eqref{eq:svd_ham}, ensuring that the energy reported by the circuit differs from the exact one only through its correlation energy. We target $\varepsilon_\infty\le 0.6$~mHa, and refer to the value of $\varepsilon_\infty(K)$ at the chosen $K$ as the compression floor.

Overall, we target an accuracy of $1.6$~mHa, corresponding to $1$~kcal/mol, in the energy that SVD-THC reports at every geometry. We divide this total error budget between two sources of error introduced by the algorithm. Data compression and bit truncation are allotted $0.6$~mHa, as imposed through Eq.~\eqref{eq:accuracy} above as a worst case over all $M$ Hamiltonians of the ensemble.
An error of $\epsilon = 1$~mHa is reserved as the resolution limit for quantum phase estimation tasks that we will introduce in Sec.~\ref{sec:results2}, setting the complexity of those algorithms to $\mathcal{N}=O(\lambda/\epsilon)$. Sliced-THC approaches are evaluated with the same error budget and are validated in the same way as SVD-THC, where each per-geometry BLISS-THC factorization is accepted at the smallest THC rank whose CCSD correlation energy error is below $0.6$~mHa. Both sides of every comparison are therefore held to the same CCSD accuracy criterion.

\subsection{Convergence and rank structure}

Figure~\ref{fig:convergence} shows the convergence of $\varepsilon_\infty(K)$ as well as the median energy error with respect to the number of retained SVD modes $K$ for the three systems. Figure~\ref{fig:convergence}$(a)$ highlights the effect of relaxation on compressibility. The rigid BCE scan contains no relaxation-induced structure and reaches $\varepsilon_\infty=2.3\times10^{-3}$~mHa by $K=9$, whereas the relaxed scan does not reach similar accuracy until $K=24$. For alanine dipeptide, $\varepsilon_\infty$ first falls below the target at $K=100$, stays between $0.53$ and $0.59$~mHa up to $K=120$, and drops to $0.22$~mHa at $K=130$. Over this plateau the maximum is attained at a single geometry, whose error is removed only once modes 121 to 130 are retained. The curve is not strictly monotone, because truncating the SVD is not variational, and adding a mode can increase the error at an individual geometry. The figure also highlights how we select $K$ for each ensemble. Our approach consists of choosing a value that clears the $0.6$~mHa target with some margin rather than at the crossing point itself. For each of these systems, we choose $K=14$, $90$, and $150$ respectively. The margin allows flexibility in the BLISS-THC compression step where more of the error budget can be spent on lowering the ranks $R_\ell$ which has a stronger effect on the total resource cost.

\begin{figure}[t!]
    \centering
    \includegraphics[width=\linewidth]{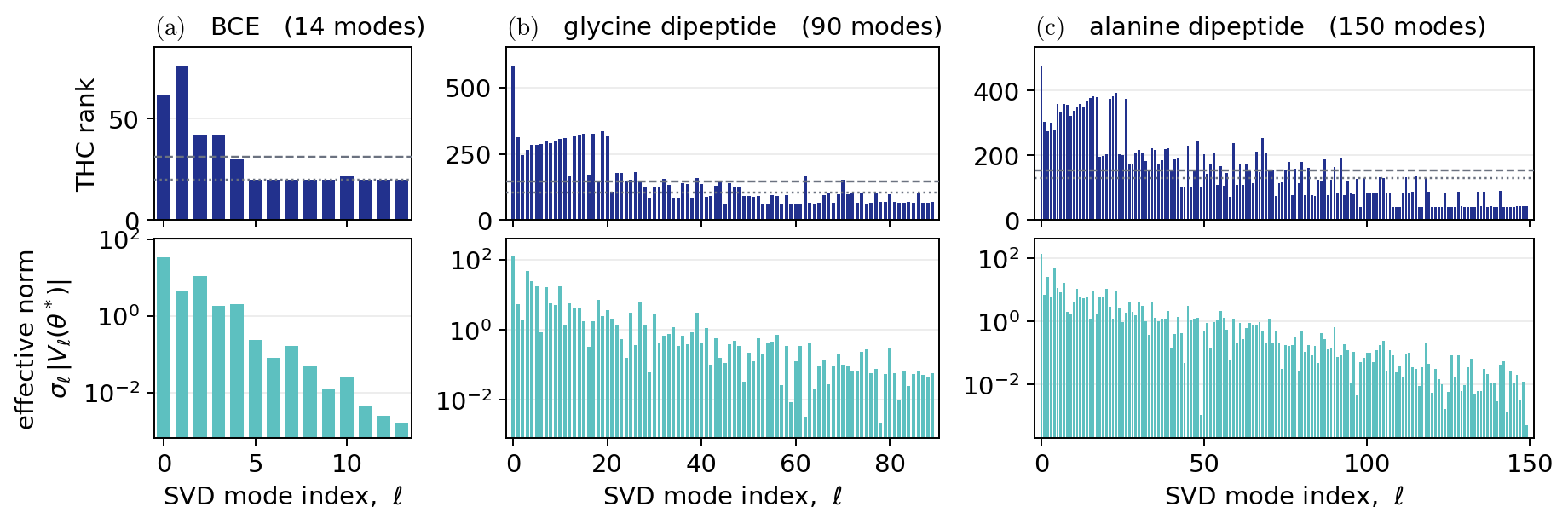}
    \caption{Per-mode structure of the three deployed SVD-THC encodings of the relaxed molecules in Table~\ref{tab:ccsd}, plotted against the SVD mode index $\ell$, with modes ordered by decreasing $\omega_\ell$. Top: THC rank $R_\ell$, with the mean (dashed) and median (dotted) rank, 31 and 20 for BCE, 146 and 104 for glycine dipeptide, and 153 and 130 for alanine dipeptide. Bottom: effective norm $\sigma_\ell |V_\ell(\boldsymbol{\theta}^*)|$ on a logarithmic scale, where $\boldsymbol{\theta}^*$ is the geometry that attains $\lambda.$ The effective norms sum exactly to $\lambda.$}
    \label{fig:permode}
\end{figure}

Figure~\ref{fig:permode} shows the two per-mode quantities that together set the cost of SVD-THC, corresponding to the THC rank $R_\ell$ at which each mode Hamiltonian is fitted, and its effective norm $\sigma_\ell|V_\ell(\boldsymbol{\theta}^*)|$ at the geometry $\boldsymbol{\theta}^*$ that attains $\lambda$, as defined in Eq.~\eqref{eq:svdthc1norm}. The modes are ordered by decreasing $\omega_\ell$, so both quantities decay with the mode index. They do not, however, decay at the same rate, and they do not induce the same ordering. We observe a right-skewed rank profile, where the mean rank exceeds the median by up to $55\%$ across the three systems, and the largest mode is fitted at two and a half to four times the average rank. This has consequences for the quantum resource estimates discussed later. Since the Clifford data-loading cost in the SVD-THC circuit scales quadratically with every rank as $O(\sum_\ell R_\ell^2)$, this short tail of high-rank modes dominates the QROAM cost rather than the number of modes $K$ itself. It is also why we allocate a different rank per mode, rather than holding it at a single common value. A uniform allocation at $\max_\ell R_\ell$ would pay the cost of the most demanding mode $K$ times over.

The effective norm decays considerably faster, falling by several orders of magnitude across the retained modes, and the three largest modes alone carry $52$--$92\%$ of $\lambda$. For both dipeptides the first mode is both the most expensive to load and the largest contributor to $\lambda$, but beyond it the orderings generally disagree. In BCE, the highest-rank mode is only the third largest in effective norm. Ultimately, both costs are front-loaded, where it is observed that half of the QROAM cost is spent on the first $14$--$16\%$ of the modes, while $90\%$ of $\lambda$ is carried by the first $21$--$24\%$.

Table~\ref{tab:ccsd} summarizes the decomposition parameters for the three SVD-THC ensembles considered: BCE at $K=14$, the glycine dipeptide at $K=90$, and the alanine dipeptide at $K=150$. The decomposition error for each ensemble is certified against Eq.~\eqref{eq:accuracy} at every geometry of its full grid, with no failures. For the sliced-THC version of the BCE molecule, block encodings for both rigid and relaxed scans obtain similar accuracy with average THC rank of $51.5$ and $51.6$ and similar 1-norms averaging $35.8$ and $36.4$ Hartree, with worst cases of $41.9$ and $50.5$ Hartree. The SVD-THC technique replaces the sliced-THC encoding scheme with the compressed circuit assembled from either $K=9$ or $K=14$ tensors. For the relaxed scan, the circuit carries a larger 1-norm than any one single-geometry block encoding, $53.8$ vs.~$50.5$ Hartree, a factor of $1.07$, and for the rigid scan $42.7$ vs.~$41.9$ Hartree, a factor of $1.02$ is obtained.

\begin{table}[t!]
\centering
\small
\setlength{\tabcolsep}{3.0pt}
\caption{Summary of the three block encodings: the torsional grid and number of geometries, the number of retained SVD modes $K$, the 1-norm $\lambda$ (Hartree), the absolute $\ell_2$ residual of the per-mode THC factorization (Hartree) summarized over the $K$ modes, and the absolute error in the CCSD correlation energy (mHa) summarized over all geometries of the grid. No geometry exceeds the target error of $0.6$~mHa for any of the three systems.}
\label{tab:ccsd}
\begin{tabular}{llccccccc}
\toprule
 & & & & & \multicolumn{2}{c}{THC residual (Ha)} &
 \multicolumn{2}{c}{CCSD $|\Delta E_{\mathrm{corr}}|$ (mHa)} \\
\cmidrule(lr){6-7}\cmidrule(lr){8-9}
\textbf{system} & \textbf{grid} & $\boldsymbol{M}$ & $\boldsymbol{K}$  & $\boldsymbol{\lambda}$ (Ha) &
\textbf{median} & \textbf{max} &
\textbf{median} & \textbf{max} \\
\midrule
1-bromo-2-chloroethane & $73\times1$ &  73 & 14 & 53.8 & 0.053 & 0.425 & 0.233 & 0.541 \\
glycine dipeptide & $37\times72$ & 2,664 & 90 & 339.9 & 0.102 & 0.517 & 0.212 & 0.589 \\
alanine dipeptide & $72\times72$ & 5,184 & 150 & 416.4 & 0.074 & 0.824 & 0.184 & 0.569 \\
\bottomrule
\end{tabular}
\end{table}

\subsection{Effective norm versus conventional norm}
\label{sec:effnorm}

The SVD-THC factorization features two types of norms.
The individual terms $\sigma_\ell|V_\ell(\boldsymbol{\theta}^*)|$ for the norm-attaining geometry $\boldsymbol{\theta}^*$ are the effective norms plotted in the lower row of Fig.~\ref{fig:permode}, and sum to $\lambda=\sum_\ell\sigma_\ell|V_\ell(\boldsymbol{\theta}^*)|$. The bare THC norms $\sigma_\ell$ that the factorization regularizes do not. The mean effective norm per mode is $3.84$~Ha for BCE, $3.78$~Ha for glycine dipeptide and $2.78$~Ha for alanine dipeptide, whereas the corresponding mean $\sigma_\ell$ is $28.2$~Ha, $164.8$~Ha and $177.7$~Ha. Only the former sums to $\lambda$, setting the true rescaling factor for the block encoding.

\begin{figure}[t]
    \centering
    \includegraphics[width=\linewidth]{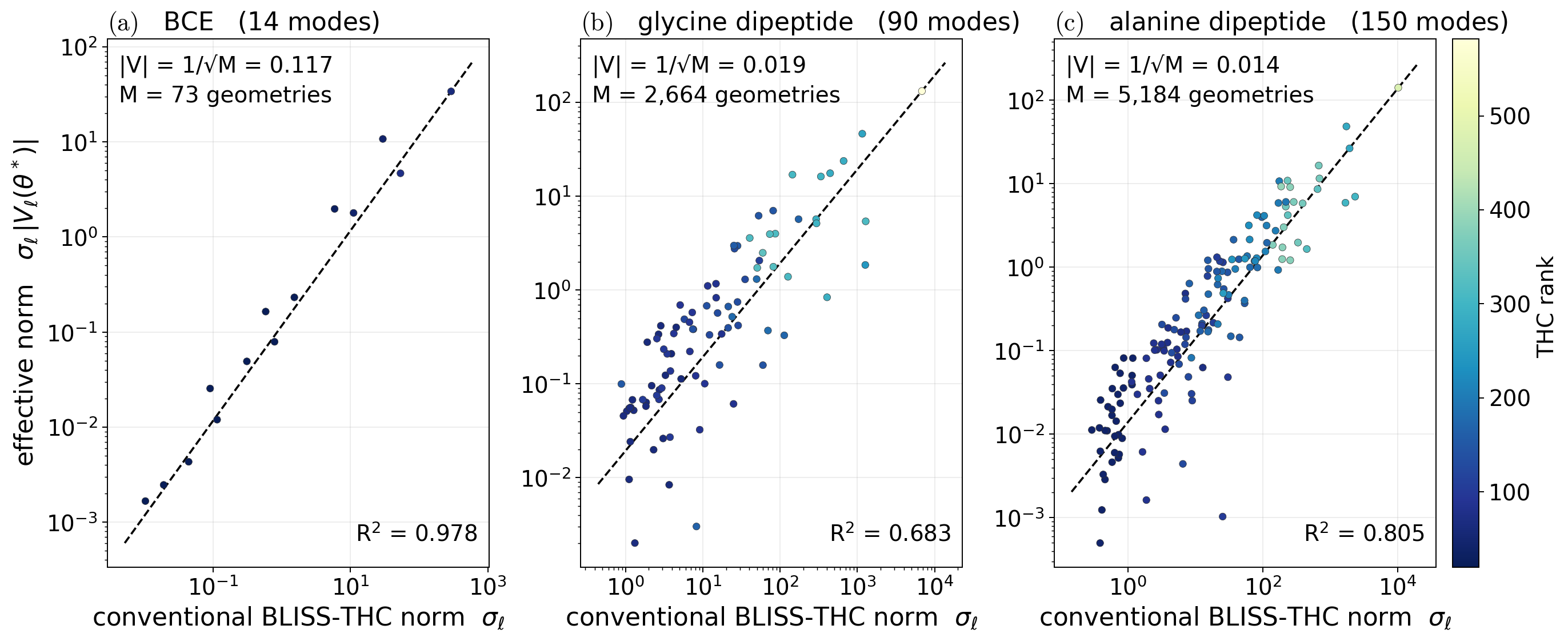}
    \caption{Effective norm $\sigma_\ell|V_\ell(\boldsymbol{\theta}^*)|$ versus the bare THC 1-norm $\sigma_\ell$ that the fit regularizes, one point per mode. The dashed line marks $|V_\ell(\boldsymbol{\theta}^*)|=1/\sqrt{M}$, the root-mean-square coefficient on an $M$-geometry grid. $R^2$ is the quality of a log-log fit of one quantity against the other.}
    \label{fig:effnorm}
\end{figure}

Applying THC to an individual tensor normally has the target to minimize its 1-norm $\sigma_\ell$. In our algorithm, the tensors are always modulated by their coordinate-dependent factor $|V_\ell(\boldsymbol{\theta}^*)|$. Figure~\ref{fig:effnorm} compares the effective norm with the bare 1-norm $\sigma_\ell$ that the THC fit regularizes. The two are only loosely correlated, and the orderings they induce diverge most where the spread of $|V_\ell(\boldsymbol{\theta}^*)|$ is widest: the dipeptides give $R^2\approx0.7$--$0.8$, while BCE remains close to rank-faithful at $R^2=0.978$. The fit therefore penalizes a quantity that is not, in general, the actual contribution to the 1-norm of the full algorithm $\lambda$ of a certain mode.

The normalization of the SVD eigenvectors determines the scale of this effect. Each column $V_\ell$ has unit norm over the $M$-point grid, so its root-mean-square value at a single geometry is $1/\sqrt{M}$, the dashed line of Fig.~\ref{fig:effnorm}. This places the dashed line for BCE, with $M=73$, almost an order of magnitude above those of the dipeptides, showing that it is a property of the grid size rather than of the molecules. More importantly, $\lambda$ is the sampled maximum of a function that is smooth in $\boldsymbol{\theta}$. Refining the grid therefore resolves the same bounded surface more finely rather than discovering unboundedly larger values. Without refitting, $\lambda$ does not change for relaxed BCE and glycine dipeptide, whose norm-attaining geometry occurs in every nested sub-grid, although their full grids are $7.3\times$ and $14.8\times$ larger than the coarsest ones. For alanine dipeptide, increasing the grid size by $576\times$ changes $\lambda$ by $24\%$, still well below the $O(M)$ cost of encoding each geometry independently. The resource estimates of Sec.~\ref{sec:qre} instead re-optimize the operator on each grid, which lowers $\lambda$ on the smaller grids (Tables~\ref{tab:bce-rigid-qre-corrected}--\ref{tab:alanine-dipeptide-qre-corrected}) without changing the query count. Thus, the contributions in Eq.~\eqref{eq:svd_ham} pay for grid resolution only once, yielding sublinear scaling with the number of geometries $M$.

\section{Quantum resource estimates}
\label{sec:results2}
In this section, we introduce an application of our new primitive for electronic structure computation on geometric ensembles, and discuss its utility in comparison to alternative quantum algorithms. To this end, we analyze the quantum resource requirements for a representative application of our SVD-THC circuit: the \emph{energy thresholding} algorithm divides an ensemble state into two subspaces, one with energy above a delimiter value $\Pi$, and one with energy below. It has potential applications in tasks such as geometry optimization and the preparation of the PES state in Eq.~\eqref{eq:pesstate}. After defining the thresholding circuit, we provide quantum resource estimates for the qubit-comparable and gate-optimal limits of the space-time trade-offs in QROAM data-loaders \cite{low2024trading}. These quantum resource estimates are compared to the baseline version of the thresholding algorithm using the sliced-THC, as well as an alternative, where each geometry is targeted with a separate and independent quantum computation.

\subsection{Energy thresholding}
\label{sec:threshold}
 
\begin{figure}
    \centering
    \includegraphics[width=0.65\linewidth]{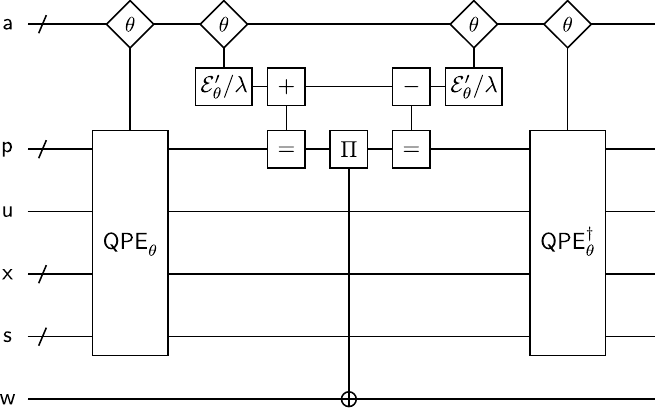}
    \caption{Thresholding circuit, using the phase estimation routine $\mathsf{QPE}_{\theta}$  and its inverse to entangle eigenstates of the Hamiltonians $\hat{H}_{\theta}$ for every angle parameter $\theta$ in register $\mathsf{a}$ with expressions for their eigenenergies in the phase register $\mathsf{p}$. $\theta$-dependent energy offsets $\mathcal{E}^\prime_{\theta}/\lambda$ are loaded or generated in a temporary register. With an in-place adder, these values correct the energy estimates in the $\mathsf{p}$ register. A comparator circuit `$\Pi$' checks the values in the $\mathsf{p}$ register and flips the flag qubit $\mathsf{w}$ if the values are below a certain energy threshold. Here, the $\mathsf{s}$ register encodes the electronic structure of the simulated system and  $\mathsf{x}$ includes all permanent auxiliary registers for the block encoding while qubit $\mathsf{u}$ is used for quantum signal processing. Diamond-shaped gates indicate multiplexed inputs but not necessarily data-loaders.}
    \label{fig:thresh}
\end{figure}

Let $|\psi_{\theta,k}\rangle$ be the $k$-th eigenstate of the Hamiltonian $\hat{H}_\theta/\lambda=\sum_k E_{\theta,k}|\psi_{\theta,k}\rangle\!\langle\psi_{\theta,k}|$ with respect to energies $E_{\theta,k}\in[-1,1]$, and let $\Pi$ be the energy threshold for the Hamiltonian $H(\theta)$, which includes the constant offset $\mathcal{E}_{\theta}$.  Using the multiplexed version of quantum phase estimation, $\mux_\theta \mathsf{QPE}_\theta$ featuring the SVD-THC circuit for $\mux_\theta \mathcal{B}[\hat{H}_\theta/\lambda]$, the thresholding circuit in Figure~\ref{fig:thresh} approximately sets a flag qubit $\mathsf{w}$ in the low-energy subspace of an ensemble state \eqref{eq:typical_state}, such that

\begin{align}
   \sum_{\theta} c_\theta 
|\theta\rangle_{\mathsf{a}}
|\phi_\theta\rangle_{\mathsf{s}} \; \mapsto  \; \sum_{\theta} c_\theta 
|\theta\rangle_{\mathsf{a}}
\sum_k \langle \psi_{\theta, k} |\phi_\theta\rangle |\psi_{\theta,k}\rangle_{\mathsf{s}} |\pi_{\theta,k}\rangle_{\mathsf{w}}
\end{align}

with

 \begin{align}
      |\pi_{\theta,k}\rangle =    \begin{cases}
        |1\rangle \; \;  \text{for} \; E_{\theta,k}+\mathcal{E}^\prime_\theta/\lambda\leq \Pi\, ,\\
        \\
        |0\rangle \; \; \text{else.}
    \end{cases}
 \end{align}

\begin{figure}
    \centering
    \includegraphics[width=\linewidth]{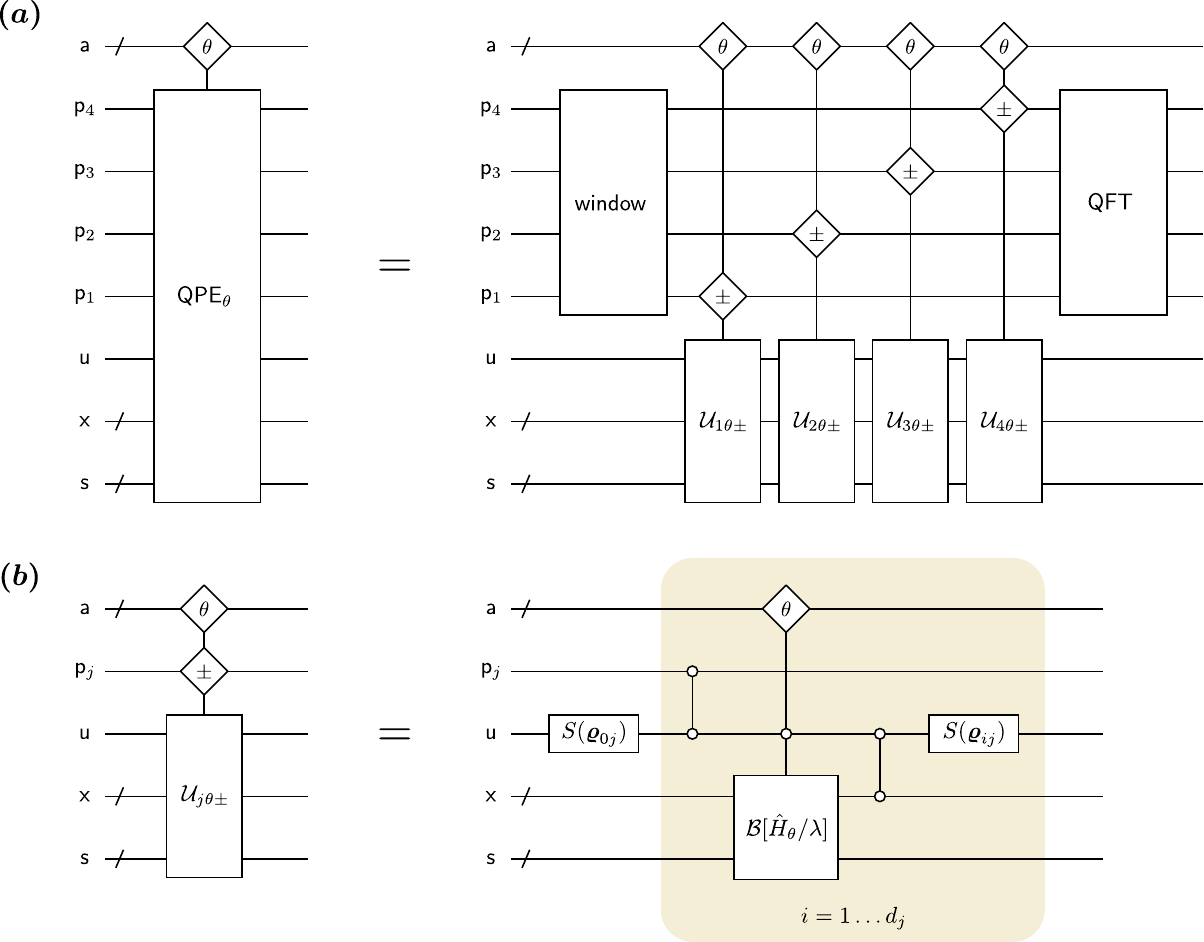}
    \caption{$\boldsymbol{(a)}$ Example of a quantum phase estimation routine with a  4-qubit phase register $\mathsf{p}$, initially prepared in a window state. Phase oracles $\mathcal{U}_{j\theta\pm}$ kick back unitary signals using bidirectional control (gates labeled `$\pm$') on phase qubits $\mathsf{p}_j$. The routine is concluded with a quantum Fourier transform $\mathsf{QFT}$. $\boldsymbol{(b)}$ Phase oracles defined by a generalized QSP routine in which $\mathsf{u}$-controlled quantum walk operators featuring the block encodings $\mathcal{B}[\hat{H}_{\theta}/\lambda]$ of $\hat{H}_{\theta}/\lambda$, as well as arbitrary 1-qubit unitaries $S(\boldsymbol{\varrho}_{ij})$, parameterized by sets of angles $\boldsymbol{\varrho}_{ij}$ for $i=0\dots d_j$ with $d_j$ being the degree of the encoded polynomial. Starting in $|0\rangle$ the qubit $\mathsf{u}$ returns to $|0\rangle$  approximately towards the end of the subroutine. The shaded region is repeated in the range of $i=1\dots d_j$. The circuit controls the sign of the Hamiltonians $\hat{H}_{\theta}$ by a controlled-$Z$ gate between qubits $\mathsf{u}$ and $\mathsf{p}_j$, allowing for double phase-kickback by implementing $\exp(i\hat{H}_{\theta}t_j/\lambda)$ in the $|0\rangle$ state of $\mathsf{p}_j$ and $\exp(-i\hat{H}_{\theta}t_j/\lambda)$ in its $|1\rangle$ state.}
    \label{fig:QPE}
\end{figure}

 Here, $\mathcal{E}^\prime_\theta$ is a correction of  $\mathcal{E}_\theta$, where we have replaced the mean-field energy of the compressed Hamiltonian by that of the exact one. Writing $E_\mathrm{HF}[\cdot]$ and $E_\mathrm{corr}[\cdot]$ for the Hartree-Fock and correlation energies of the operator part of a Hamiltonian, that is, excluding its scalar $\mathcal{E}_\theta$, we load the offset register with 
\begin{align}
    \mathcal{E}'_\theta = \mathcal{E}_{{\theta}} + E_\mathrm{HF} [H(\theta)] - E_\mathrm{HF} [{H}_\mathrm{SVD}(\theta)]\, ,
    \label{eq:hf_offset}
\end{align}
for $H(\theta)$ and $H_{\mathrm{SVD}}(\theta)$ defined in Eq.~\eqref{eq:ham} and Eq.~\eqref{eq:svd_ham}. For the ground state ($k=0$), the value $\lambda E_{\theta,0}+\mathcal{E}_{\theta}^\prime$ thus describes molecular energies,
\begin{align}
\mathcal{E}_{{\theta}} + E_\mathrm{HF} [H(\theta)]  + E_\mathrm{corr} [H_{\mathrm{SVD}}(\theta)],
\end{align}
such that the approximate mean-field energy is removed, and only the error of the correlation energy remains. It is ultimately this step that makes the correlation energy error of Sec.~\ref{sec:accuracy} the right accuracy metric, with CCSD serving as a classical proxy for the correlation energy that phase estimation resolves. The correction of the energies $E_{\theta, k}$ with a scalar $\mathcal{E}^\prime_\theta/\lambda$ requires us to compute the eigenvalues of $\hat{H}_\theta/\lambda$ dynamically. To that end, $\mux_\theta \mathsf{QPE}_\theta$,  writes estimates for the energies ${E_{\theta, k}}$ into a phase register $\mathsf{p}$:
\begin{align}
   \sum_{\theta} c_\theta 
|\theta\rangle_{\mathsf{a}}
|\phi_\theta\rangle_{\mathsf{s}} \; \mapsto  \; \sum_{\theta} c_\theta 
|\theta\rangle_{\mathsf{a}}
\sum_k \langle \psi_{\theta, k} |\phi_\theta\rangle |\psi_{\theta,k}\rangle_{\mathsf{s}} |\widetilde{E}_{\theta,k}\rangle_{\mathsf{p}}\, .
\end{align}

While $|\widetilde{E}_{\theta,k}\rangle$ are not necessarily computational basis states due to the discretization error of $\mathsf{QPE}_\theta$,  we may assume that $|\widetilde{E}_{\theta,k}\rangle$ are superpositions of finite-bit energy values $|E\rangle$ close to $E_{\theta,k}$. Through an in-place adder, the circuit in Figure~\ref{fig:thresh} corrects these energy values $E$ by the offsets  $\mathcal{E}^\prime_{\theta}/\lambda$ loaded into a temporary register, before setting the flag qubit $\mathsf{w}$. Corrections and QPE are then reversed in an attempt  to uncompute the phase register $\mathsf{p}$, which will succeed only approximately.

Note that we expect this correction with values $\mathcal{E}^\prime_\theta/\lambda$ to be necessary for many other applications, but a regular, qubitization-based QPE circuit would (approximately) write expressions such as $|\arccos E_{\theta,k}\rangle$ rather than energies $|E_{\theta, k}\rangle$ into its phase register, making additive corrections impossible.
Algorithms such as the energy thresholding circuit therefore need to run on quantum signal processing, rather than qubitization. This means that the phase kickback operators $\mathcal{U}_{j\theta\pm}$ within the multiplexed QPE circuit, depicted in detail in Figure~\ref{fig:QPE}$(a)$, must encode time-evolution operators $\exp(\pm i\hat{H}_{\theta}t_j/\lambda)$ rather than qubitization eigenvalues $\exp(\pm i t_j\arccos E_{\theta,k})$. To turn the latter into the former, we employ a generalized quantum signal processing (GQSP)~\cite{motlagh2024generalized} circuit,  depicted in Figure~\ref{fig:QPE}$(b)$, that repeatedly queries the SVD-THC circuit $\mux_\theta \mathcal{B}[\hat{H}_\theta/\lambda]$. For phase kickback and signal modulation, GQSP introduces a qubit $\mathsf{u}$. For every phase-kickback operator $\mathcal{U}_{j\theta\pm}$, we define a GQSP circuit that uses the Jacobi-Anger expansion polynomial of Ref.~\cite{low2019hamiltonian} with degree $d_j = O(t_j+\log\varepsilon^{-1}_j)$, to approximate
\begin{align}
\mathcal{U}_{j\theta\pm}\approx e^{\pm i\hat{H}_{\theta}t_j/\lambda} \otimes |0\rangle\!\langle0|_{\mathsf{u}} \otimes  |0\rangle\!\langle0|_{\mathsf{x}}+ ...
\end{align}
up to an error of $\varepsilon_{j}$ in the operator distance. Here the register $\mathsf{x}$ is the amalgamation of all permanent auxiliary registers used in the block encoding of Figure~\ref{fig:blockencs}:  $\mathsf{x}=\mathsf{bg}_3\mathsf{j}_5\mathsf{vq}_{\lbrace 0,3,4,5,7,8 \rbrace}$ and $t_j$ is a time parameter large enough to estimate the $j$-th bits in values for $\hat{H}_\theta/\lambda$.  In an $n_{\mathrm{p}}$-qubit QPE routine, we implement a series of multiplexed subroutines $\mathcal{U}_{j\theta\pm}$ for $j=1\dots n_{\mathrm{p}}$, in which the phase qubits $\mathsf{p}_j$ control the direction $\pm$ of the time evolution operators (double phase-kickback). For every such phase oracle we make the choice $t_j=2^{j-1}\pi$ and $\varepsilon_j =  \varepsilon 2^{-(n_{\mathrm{p}}+1-j)} $ for some error $\varepsilon$. In that way, our QPE queries the SVD-THC circuit a total of 
\begin{align}
    \sum_{j=1}^{n_{\mathrm{p}}} d_j = O\!\left(2^{n_{\mathrm{p}}} + n_{\mathrm{p}}^2 + n_{\mathrm{p}}\log\varepsilon^{-1}\right)
\end{align}
times. Due to mixing different approximations to the time evolution inside the QPE circuit, we have a somewhat lower confidence in the estimated phase values. Let us say that in a comparable version of $\mux_{\theta} \mathsf{QPE}_{\theta}$, where the phase kickback operators $\mathcal{U}_{j\theta\pm}$ are replaced by the exact time evolution operators $\exp(\pm i\hat{H}_{\theta}t_j/\lambda)$, the probability of a phase estimate $\phi^\prime$ being at least $\Delta$-close to the actual phase $\phi$ is $\eta$, i.e., $\mathrm{Pr}(|\phi-\phi^\prime|\leq \Delta) = \eta$. Then, in the QPE of Figure~\ref{fig:QPE}, we have an updated confidence window for phase estimates $\phi^{\diamond}$:

\begin{align}
    \mathrm{Pr}(|\phi^{\diamond}-\phi|\leq \Delta)\;&  \geq\; \eta-\frac{1}{2}\sum_{j}\varepsilon_j \\ &\geq  \;  \eta - \varepsilon /2\, ,
\end{align}
as is illustrated in Appendix~\ref{sec:newpe}. By defining the approximation errors $\varepsilon_j = \varepsilon2^{-(n_{\mathrm{p}}+1-j)}$ to become smaller with smaller values of $j$, we have reduced the uncertainty on the more significant bits. For QPE circuits in this paper, we use the convention $\varepsilon = 0.1$, which corresponds to a $5\%$ drop in confidence.

\FloatBarrier
\subsection{Cost analysis}
\label{sec:qre}

Based on the energy threshold algorithm, we now provide quantum resource estimates for different grid sizes of the three molecular benchmark systems presented in Sec.~\ref{sec:results}. Specifically, we want to compare the SVD-THC version of the algorithm to the sliced-THC version, in which each geometric subspace is assigned a Hamiltonian via slicing of single-geometry THC circuits. There, each single-geometry block encoding is a BLISS-THC factorization fitted without correlation-energy reweighting, at the smallest rank whose CCSD correlation-energy error is below $0.6$~mHa, the same CCSD validation applied to SVD-THC (Sec.~\ref{sec:accuracy}). For SVD-THC, the multiplexed QPE  has a Jacobi-Anger schedule with binary time steps, so the cost scales with the power-of-two integer above $\lambda/\epsilon$ rather than smoothly with $\lambda/\epsilon$ itself. A non-integer-power-of-two QPE using Jacobi-Anger expansion is beyond the scope of this work. This introduces a rung-dependent overhead between 1$\times$ and 2$\times$, and because the sliced-THC and SVD-THC Hamiltonians can land on different rungs, their apparent speed-up can shift. To somewhat mitigate this effect, reported comparisons use the same Jacobi-Anger counting procedure in both SVD-THC and sliced-THC QPEs. For QPE query complexity $\mathcal{N}$, we use the convention $\mathcal{N} \sim \lceil \pi \lambda/\epsilon \rceil$ based on single-shot statistical confidence, whereas the alternative Holevo-based normalization would use $\mathcal{N} \sim \lceil \pi \lambda/(2\epsilon)\rceil $. This is an overall factor-of-two deviation from the quantum phase estimation normalization, independent of the Jacobi-Anger scheduling. In the results below, both SVD-THC and sliced-THC estimates are evaluated under the same convention. 
\begin{figure}[t!]
\centering
\includegraphics[width=0.65\linewidth]{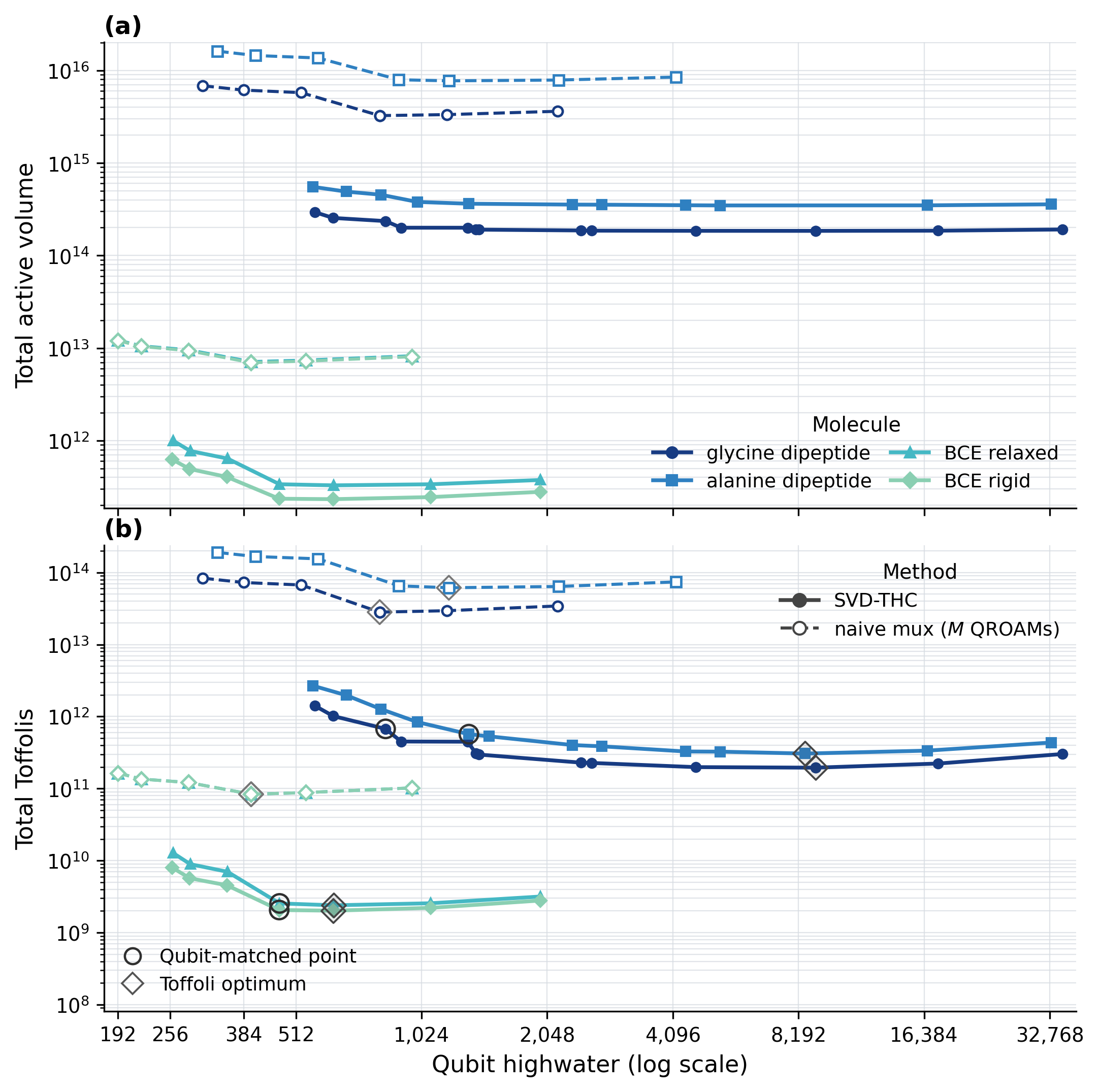}
\caption{Space-time trade-off for the sliced-THC (open markers, dashed) and SVD-THC version (filled markers, solid) of the thresholding circuit, sweeping the qubit highwater against $\boldsymbol{(a)}$ the total active volume and $\boldsymbol{(b)}$ total Toffoli counts. Each curve is one molecule at its full grid. Outlined diamonds mark each method's Toffoli minimum point. Outlined circles mark the qubit-comparable point of SVD-THC, defined relative to the Toffoli-optimal point of the sliced-THC multiplexed scheme.}
\label{fig:tradeoffs}
\end{figure}

Figure~\ref{fig:tradeoffs} summarizes the space-time trade-offs for SVD-THC across all molecules. Each point is obtained by independently varying the QROAM qubit workspace allocation as well as the Givens angle-loading batching strategy. By varying both the number of angle batches and the QROAM space-time trade-off parameter, we obtain a Pareto curve for both active volume and total Toffoli count. Within these two plots, we observe three features which organize the rest of the section. First, the SVD-THC curves lie below the sliced-THC curves across the whole qubit range. Second, the Toffoli count falls steadily and gives way to a genuine minimum (highlighted by the diamond markers), and rises again once the swap network associated with the QROAM is overprovisioned. Third, the active volume flattens much earlier compared to Toffoli count. For the dipeptides, it comes within $3\%$ of its minimum by about 1400 qubits, and it is essentially flat beyond that. The data-loading cost that dominates the inner $\mathsf{prep}$ is set by the total size of the table, rather than how it is partitioned; so past this point the additional qubits only reduce Toffolis, with little effect on the overall active volume. 

We therefore report two operating points on these curves. The \emph{qubit comparable} setting is defined as the closest approximation of the SVD-THC version of the circuit to the gate-optimal version of the sliced-THC circuit in terms of qubits. The \emph{Toffoli optimal} setting is the minimum of the Toffoli count, which sometimes coincides with the true minimum of the active volume curve.

Table~\ref{tab:bce-rigid-qre-corrected} and Table~\ref{tab:bce-relaxed-qre-corrected} summarize the quantum resource estimates for both rigid and relaxed BCE. At the coarsest grid of Table~\ref{tab:bce-relaxed-qre-corrected}, which contains ten geometries, the SVD-THC requires $4.7\times$ fewer Toffoli gates at the qubit-comparable setting, or 5.0$\times$ at the Toffoli optimal one. Over the full 73-point grid, the overall advantage reaches 32.8$\times$ at the qubit-comparable setting, and $34.9\times$ at the Toffoli-optimal one. As expected, the rigid scan shows larger compressibility and  provides larger improvements overall. The same pattern holds for the dipeptides, where the SVD-THC 1-norm $\lambda$ exceeds the maximal single-geometry 1-norm $\lambda_{\mathrm{max}}$ by a factor of 2.5 and 2.7, yet the Toffoli advantage reaches $41.8\times$ and $108.0\times$ at the qubit-comparable setting, and $144.5\times$ and $200.6\times$ at the Toffoli-optimal setting, for glycine dipeptide and alanine dipeptide respectively.

\begin{table*}[t]
\centering
\small
\setlength{\tabcolsep}{2.6pt}
\setlength{\aboverulesep}{0pt}\setlength{\belowrulesep}{0pt}
\renewcommand{\arraystretch}{1.3}
\begin{tabular*}{\textwidth}{@{\extracolsep{\fill}}c|ccc|cccc|cccc}
\toprule
& \multicolumn{3}{c|}{Sliced-THC} & \multicolumn{4}{c|}{SVD-THC} & \multicolumn{4}{c}{SVD-THC} \\
& \multicolumn{3}{c|}{} & \multicolumn{4}{c|}{{qubit comparable}} & \multicolumn{4}{c}{{Toffoli optimal}} \\
\cmidrule{2-12}
$M$ & Qubits & AV & Toff. & Qubits & AV & Toff. & Impr. & Qubits & AV & Toff. & Impr. \\[-2pt]
 &  & {\footnotesize$(10^{11})$} & {\footnotesize$(10^{10})$} &  & {\footnotesize$(10^{11})$} & {\footnotesize$(10^{9})$} &  &  & {\footnotesize$(10^{11})$} & {\footnotesize$(10^{9})$} &  \\
\midrule
10 & 397 & 9.52 & 1.14 & 458 & 2.28 & 1.97 & $\mathbf{5.8\times}$ & 620 & 2.25 & 1.92 & $\mathbf{6.0\times}$ \\
19 & 398 & 18.1 & 2.17 & 461 & 2.30 & 2.00 & $\mathbf{10.9\times}$ & 623 & 2.27 & 1.96 & $\mathbf{11.1\times}$ \\
37 & 399 & 35.2 & 4.23 & 464 & 2.32 & 2.03 & $\mathbf{20.9\times}$ & 626 & 2.29 & 1.98 & $\mathbf{21.4\times}$ \\
55 & 399 & 52.4 & 6.29 & 465 & 2.36 & 2.07 & $\mathbf{30.4\times}$ & 627 & 2.33 & 2.03 & $\mathbf{31.0\times}$ \\
73 & 400 & 69.5 & 8.35 & 467 & 2.35 & 2.06 & $\mathbf{40.5\times}$ & 629 & 2.33 & 2.02 & $\mathbf{41.4\times}$ \\
\bottomrule
\end{tabular*}
\caption{Resource estimates  of the energy thresholding circuit for the rigid BCE torsion scan of different grid sizes containing qubit, Toffoli and AV counts for versions of the algorithm featuring sliced-THC  as well as the qubit-comparable and Toffoli-optimal SVD-THC. `{Impr.}' is the ratio of the Toffoli counts of sliced-THC and SVD-THC versions of the thresholding circuit. Active volumes and Toffolis are given in the units shown below each column heading. The qubit-comparable setting is the narrowest SVD-THC circuit whose qubit highwater exceeds that of the sliced-THC circuit at its Toffoli minimum, and the Toffoli-optimal setting is the minimum of the Toffoli count, highlighted in Fig.~\ref{fig:tradeoffs}. In the sliced-THC circuits the QROAM workspace is reused. SVD-THC compression is redone for each grid, with $\lambda=42.7$~Ha on the full grid and $\lambda=40.4$--$41.7$~Ha on the smaller grids.} 
\label{tab:bce-rigid-qre-corrected}
\par\noindent
\end{table*}
\begin{table*}[ht!]
\centering
\small
\setlength{\tabcolsep}{2.6pt}
\setlength{\aboverulesep}{0pt}\setlength{\belowrulesep}{0pt}
\renewcommand{\arraystretch}{1.3}
\begin{tabular*}{\textwidth}{@{\extracolsep{\fill}}c|ccc|cccc|cccc}
\toprule
& \multicolumn{3}{c|}{Sliced-THC} & \multicolumn{4}{c|}{SVD-THC} & \multicolumn{4}{c}{SVD-THC} \\
& \multicolumn{3}{c|}{} & \multicolumn{4}{c|}{{qubit comparable}} & \multicolumn{4}{c}{{Toffoli optimal}} \\
\cmidrule{2-12}

$M$ & Qubits & AV & Toff. & Qubits & AV & Toff. & Impr. & Qubits & AV & Toff. & Impr. \\[-2pt]
 &  & {\footnotesize$(10^{11})$} & {\footnotesize$(10^{10})$} &  & {\footnotesize$(10^{11})$} & {\footnotesize$(10^{9})$} &  &  & {\footnotesize$(10^{11})$} & {\footnotesize$(10^{9})$} &  \\
\midrule
10 & 397 & 9.78 & 1.14 & 459 & 3.26 & 2.44 & $\mathbf{4.7\times}$ & 622 & 3.17 & 2.29 & $\mathbf{5.0\times}$ \\
19 & 398 & 18.6 & 2.17 & 462 & 3.30 & 2.47 & $\mathbf{8.8\times}$ & 625 & 3.21 & 2.32 & $\mathbf{9.4\times}$ \\
37 & 399 & 36.2 & 4.24 & 465 & 3.33 & 2.51 & $\mathbf{16.9\times}$ & 628 & 3.24 & 2.36 & $\mathbf{18.0\times}$ \\
55 & 399 & 53.8 & 6.30 & 465 & 3.36 & 2.55 & $\mathbf{24.7\times}$ & 628 & 3.27 & 2.40 & $\mathbf{26.3\times}$ \\
73 & 400 & 71.4 & 8.36 & 468 & 3.38 & 2.55 & $\mathbf{32.8\times}$ & 631 & 3.29 & 2.40 & $\mathbf{34.9\times}$ \\
\bottomrule
\end{tabular*}
\caption{Resource estimates (qubit, Toffoli and AV counts) of the energy thresholding circuit for the relaxed BCE torsion scan in different grid sizes, with relative improvements between sliced-THC and SVD-THC versions of the algorithm in terms of Toffoli gates. Columns and settings as in Table~\ref{tab:bce-rigid-qre-corrected}; $\lambda=53.8$~Ha on the full grid and $53.2$--$53.8$~Ha on the smaller grids.} 
\label{tab:bce-relaxed-qre-corrected}
\end{table*}

\begin{table*}[ht!]
\centering
\small
\setlength{\tabcolsep}{2.6pt}
\setlength{\aboverulesep}{0pt}\setlength{\belowrulesep}{0pt}
\renewcommand{\arraystretch}{1.3}
\begin{tabular}{cc|ccc|cccc|cccc}
\toprule
& & \multicolumn{3}{c|}{Sliced-THC} & \multicolumn{4}{c|}{SVD-THC} & \multicolumn{4}{c}{SVD-THC} \\
& & \multicolumn{3}{c|}{} & \multicolumn{4}{c|}{{qubit comparable}} & \multicolumn{4}{c}{{Toffoli optimal}} \\
\cmidrule{3-13}
Grid & $M$ & Qubits & AV & Toff. & Qubits & AV & Toff. & Impr. & Qubits & AV & Toff. & Impr. \\[-2pt]
 &  &  & {\footnotesize$(10^{14})$} & {\footnotesize$(10^{12})$} &  & {\footnotesize$(10^{14})$} & {\footnotesize$(10^{11})$} &  &  & {\footnotesize$(10^{14})$} & {\footnotesize$(10^{11})$} &  \\
\midrule
10$\times$18 & 180 & 809 & 2.20 & 1.91 & 828 & 2.24 & 6.62 & $\mathbf{2.9\times}$ & 8{,}988 & 1.73 & 1.82 & $\mathbf{10.5\times}$ \\
19$\times$18 & 342 & 810 & 4.17 & 3.62 & 831 & 2.24 & 6.64 & $\mathbf{5.5\times}$ & 8{,}991 & 1.74 & 1.84 & $\mathbf{19.7\times}$ \\
19$\times$36 & 684 & 811 & 8.35 & 7.24 & 834 & 2.26 & 6.66 & $\mathbf{10.9\times}$ & 8{,}994 & 1.75 & 1.86 & $\mathbf{38.8\times}$ \\
37$\times$36 & 1{,}332 & 812 & 16.3 & 14.1 & 837 & 2.29 & 6.70 & $\mathbf{21.0\times}$ & 8{,}997 & 1.78 & 1.90 & $\mathbf{74.1\times}$ \\
37$\times$72 & 2{,}664 & 813 & 32.5 & 28.2 & 840 & 2.36 & 6.75 & $\mathbf{41.8\times}$ & 9{,}000 & 1.85 & 1.95 & $\mathbf{144.5\times}$ \\
\bottomrule
\end{tabular}
\caption{Resource estimates of the energy thresholding circuit for the glycine dipeptide in different grid sizes, including relative improvements of the sliced-THC version and SVD-THC versions of the algorithm in terms of Toffoli gates. Columns and settings as in Table~\ref{tab:bce-rigid-qre-corrected}; $\lambda=339.9$~Ha on the full grid and $321.2$--$337.3$~Ha on the smaller grids.} 
\label{tab:glycine-qre-corrected}
\end{table*}

\begin{table*}[t]
\centering
\small
\setlength{\tabcolsep}{2.6pt}
\setlength{\aboverulesep}{0pt}\setlength{\belowrulesep}{0pt}
\renewcommand{\arraystretch}{1.3}
\begin{tabular}{cc|ccc|cccc|cccc}
\toprule
& & \multicolumn{3}{c|}{Sliced-THC} & \multicolumn{4}{c|}{SVD-THC} & \multicolumn{4}{c}{SVD-THC} \\
& & \multicolumn{3}{c|}{} & \multicolumn{4}{c|}{{qubit comparable}} & \multicolumn{4}{c}{{Toffoli optimal}} \\
\cmidrule{3-13}
Grid & $M$ & Qubits & AV & Toff. & Qubits & AV & Toff. & Impr. & Qubits & AV & Toff. & Impr. \\[-2pt]
 &  &  & {\footnotesize$(10^{13})$} & {\footnotesize$(10^{11})$} &  & {\footnotesize$(10^{14})$} & {\footnotesize$(10^{11})$} &  &  & {\footnotesize$(10^{14})$} & {\footnotesize$(10^{11})$} &  \\
\midrule
3$\times$3 & 9 & 1{,}182 & 1.34 & 1.07 & 1{,}302 & 3.21 & 5.38 & $\mathbf{0.20\times}$ & 8{,}470 & 3.08 & 2.75 & $\mathbf{0.39\times}$ \\
9$\times$9 & 81 & 1{,}185 & 12.1 & 9.61 & 1{,}312 & 3.20 & 5.41 & $\mathbf{1.8\times}$ & 8{,}480 & 3.07 & 2.78 & $\mathbf{3.5\times}$ \\
18$\times$18 & 324 & 1{,}187 & 48.3 & 38.4 & 1{,}318 & 3.27 & 5.45 & $\mathbf{7.0\times}$ & 8{,}486 & 3.14 & 2.82 & $\mathbf{13.6\times}$ \\
36$\times$36 & 1{,}296 & 1{,}189 & 193 & 154 & 1{,}324 & 3.35 & 5.53 & $\mathbf{27.8\times}$ & 8{,}492 & 3.22 & 2.90 & $\mathbf{53.0\times}$ \\
72$\times$72 & 5{,}184 & 1{,}191 & 772 & 615 & 1{,}330 & 3.63 & 5.70 & $\mathbf{108.0\times}$ & 8{,}498 & 3.50 & 3.06 & $\mathbf{200.6\times}$ \\
\bottomrule
\end{tabular}
\caption{Resource estimates of the thresholding circuit for  alanine dipeptide in different grid sizes, including relative improvements between sliced-THC and two different SVD-THC versions of the algorithm. Columns and settings as in Table~\ref{tab:bce-rigid-qre-corrected}; $\lambda=416.4$~Ha on the full grid and $268.7$--$376.5$~Ha on the smaller grids.} 
\label{tab:alanine-dipeptide-qre-corrected}
\end{table*}

As established in Sec.~\ref{sec:effnorm}, the number of retained modes $K$, the ranks $R_\ell$, and $\lambda$ are properties of the SVD-THC technique rather than of the grid. Both the QROAM cost and $\lambda$ vary only weakly as a function of grid size, so the SVD-THC Toffoli count remains within $6\%$ at the qubit-comparable setting and $12\%$ at the Toffoli-optimal one of its coarsest-grid value, while the counts for the sliced-THC version of the algorithm add the resources of the block encodings of every geometry and so grow linearly with the grid size. The improvement over the baseline circuit therefore grows with $M$, at the qubit-comparable setting from $4.7\times$ to $32.8\times$ for BCE between $10$ and $73$ geometries, from $2.9\times$ to $41.8\times$ for glycine dipeptide between 180 and 2,664 geometries, and from $0.20\times$ to $108.0\times$ for alanine dipeptide between 9 and 5,184 geometries, and at the Toffoli-optimal setting from $5.0\times$ to $34.9\times$ (rigid BCE: $6.0\times$ to $41.4\times$), $10.5\times$ to $144.5\times$ and $0.39\times$ to $200.6\times$ respectively.

Furthermore, we observe an improvement for alanine dipeptide below unity at the beginning. At the coarsest $3\times 3$ grid, the fixed cost of a circuit sized for the full 150-mode ensemble, whose largest rank is 477, is not yet amortized, and SVD-THC is more expensive than sliced-THC. The advantage becomes positive with enough geometries, which occurs between 9 and 81 geometries at both settings, growing thereafter. This is the circuit-level counterpart of the behavior of $\lambda$ discussed in Sec.~\ref{sec:effnorm}. Compression becomes profitable once the grid is dense enough that redundant per-geometry data-loading, rather than the shared ensemble, dominates the cost.

\begin{figure}[t!]
\centering
\includegraphics[width=\linewidth]{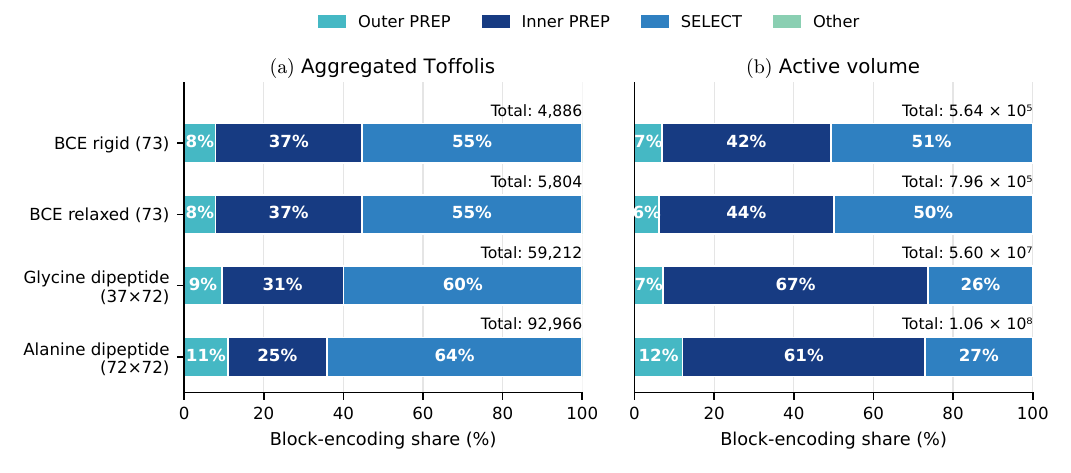}
\vspace{-0.2cm}
\caption{Breakdown of $\boldsymbol{(a)}$ aggregated Toffolis and $\boldsymbol{(b)}$ active volume for a single SVD-THC block encoding query at the full-grid Toffoli-optimal operating point, split into subroutines $\mux_{\theta} \mathsf{prep}_{\mathsf{o},\theta}$, $\mux_{\ell} \mathsf{prep}_{\mathsf{i},\ell}$ and $\mux_\ell \mathsf{Select}_\ell$ from Fig.~\ref{fig:blockencs}, labeled `Outer PREP', `Inner PREP' and `SELECT' in this figure. Other contributions are below $1\%$ for all systems and cost types and are not visible at this scale. The total above each bar is the cost of one query.}
\label{fig:subroutines}
\end{figure}

The improvements observed in the active volume metric are smaller. At the Toffoli-optimal setting the full-grid active-volume improvements are 29.9$\times$ and 21.7$\times$ for rigid and relaxed BCE, 17.6$\times$ for glycine dipeptide and 22.1$\times$ for alanine dipeptide, against Toffoli improvements of 41.4$\times$, 34.9$\times$, 144.5$\times$ and 200.6$\times$, so for the dipeptides close to $90\%$ of the Toffoli advantage does not carry over into active volume. To understand why, Fig.~\ref{fig:subroutines} divides one SVD-THC block encoding query, at the full-grid Toffoli-optimal operating point, into its subroutines: the outer $\mathsf{prep}$ ($\mux_\theta \mathsf{prep}_{\mathsf{o},\theta}$), which loads the geometry-dependent amplitudes $\sigma_\ell V_\ell(\theta)/\lambda$ and is the only part of the circuit that addresses the geometry register; the inner $\mathsf{prep}$ ($\mux_\ell \mathsf{prep}_{\mathsf{i},\ell}$),  which loads amplitudes of the central tensors of the modes set by the outer $\mathsf{prep}$; and  $\mathsf{Select}$ ($\mux_\ell \mathsf{Select}_\ell$), which applies that mode’s basis rotations and fermionic operators. We observe two features. First, the outer $\mathsf{prep}$ accounts for only a small share of the cost, carrying $8$--$11\%$ of the Toffolis and $6$--$12\%$ of the active volume, so the one subroutine whose data grows with the number of geometries is also a small share of every query, which is the structural reason the SVD-THC costs of Tables~\ref{tab:bce-rigid-qre-corrected}-\ref{tab:alanine-dipeptide-qre-corrected} are nearly flat in grid size. Second, the two cost measures attribute the remainder differently. In terms of Toffolis,  $\mathsf{Select}$ dominates for every system at $55$--$64\%$ of the total cost. In active volume the dipeptides instead place $61$--$67\%$ of the cost in the inner $\mathsf{prep}$, while for BCE, whose tensors are an order of magnitude smaller,  $\mathsf{Select}$ still leads at about $50\%$. The inner $\mathsf{prep}$ consists of a QROAM read of the central tensors, and active volume, unlike the Toffoli count, also prices the Clifford operations. These Cliffords are the dominant contribution when writing the loaded data and their number grows with the size of the table. It is this data-loading that negates most of the Toffoli advantage when counting active volume blocks.

Since the sliced-THC cost is linear in the number of geometries while the SVD-THC cost is nearly constant, every improvement factor is close to proportional to $M$, and the break-even grid size follows directly. In terms of Toffolis, the dipeptides break-even at about $17$--$23$ geometries at the Toffoli-optimal setting and $45$--$65$ at the qubit-comparable one; in active volume, the break-even moves out to $140$--$210$ geometries. These results are  qualitatively similar to protocols where instead of a thresholding circuit, an energy estimation is done for each geometry using BLISS-THC. We discuss concrete resource counts in Appendix~\ref{sec:sequential}.

\section{Conclusion}
\label{sec:conclusion}
In this work, we have developed a novel framework for electronic structure calculations across multiple geometries of a molecule, encoded in superposition on a quantum computer. In particular, we have shown that the electronic system of every geometry can be addressed simultaneously beyond the break-even point with methods that treat them separately.
With SVD-THC we have provided the key primitive to enable a shared time evolution of all geometric subspaces, allowing the parallel computation of molecular energies in quantum phase estimation.
Analyzing the quantum resource counts of SVD-THC circuits  across several molecular systems and grid sizes, we achieve a positive compression beyond the aforementioned break-even point for grids above tens of geometries.

On the largest grids considered, we obtain Toffoli reductions of $35\times$ to $41\times$ for BCE, $144 \times$ for glycine dipeptide, and $200\times$ for alanine dipeptide. These improvements grow almost in proportion to the number of geometries $M$, demonstrating that the cost of representing the ensemble can become substantially smaller than repeating an independent computation for every geometry. The advantage persists when measured in active volume, although it is reduced to $18\times$ to $30\times$. This difference highlights the cost of the Clifford-heavy data-loading operations that are largely invisible to the Toffoli metric. Reducing this data-loading overhead is therefore an important direction for future work. In particular, techniques for loading compressed representations of the geometry-dependent data could substantially narrow the gap between Toffoli count and active volume. Achieving comparable compression in active volume, even with a modest increase in qubit count, would provide a stronger indication of the practical utility of SVD-THC.

We hope this work presents a useful new primitive that will be a part of algorithms widening the range of scientific questions that can be answered with a quantum computer. For instance the Liouvillian framework for performing Born-Oppenheimer molecular dynamics~\cite{Simon2024Improved} relies on Hamiltonian simulation controlled on a grid over the nuclear positions. Similarly, quantum-search algorithms~\cite{Grover1996Fast,Durr1996Quantum} could make use of SVD-THC when constructing oracles based on molecular energies, enabling searches for minimum-energy geometries, barrier heights, transition probabilities, or just sampled geometries with an energy below a certain threshold.  For instance, an unstructured search for the lowest-energy geometry requires only $O(\sqrt{M})$ applications of the thresholding circuit, whereas direct estimation requires a separate phase estimation for each of the $M$ geometries. An SVD-THC-powered search therefore becomes favorable once the compression advantage over sequential QPE (Appendix~\ref{sec:sequential}) exceeds $\sqrt{M}$. This is already the case for BCE on its full grid, while for the dipeptides the advantage is comparable to $\sqrt{M}$ at the grid sizes considered here, so the outcome depends on the constant factors of the search. Since the advantage grows roughly in proportion to $M$, a clear advantage is expected on moderately larger grids.

The framework is not restricted to grids over internal molecular coordinates or to SVD-THC as the underlying tensor factorization. Other low-rank representations of the electron-repulsion integrals, such as sum-of-squares decompositions, could provide alternative routes for compressing the geometry-dependent Hamiltonians and may offer different trade-offs between classical preprocessing, qubit count, and circuit complexity. Furthermore, our method may be applicable to other types of  Hamiltonian ensembles. A particularly relevant example is a conformer ensemble, i.e., a set of low-energy geometries, as generated for instance by classical conformer search methods. In this case, the geometries need not lie on a regular grid. Distinct conformers typically differ by larger and more global rearrangements than neighboring grid points. We might therefore expect the shared low-rank structure exploited by our compression to be less prominent. On the other hand, all conformers share the same atoms, bonding topology and much of their local electronic structure, which is the redundancy that SVD-THC exploits. Quantifying how well the compression works with conformer ensembles is therefore an important next step, in particular for computational drug discovery, where the bioactive conformation of a ligand, its strain energy upon binding, and Boltzmann-weighted molecular properties all hinge on an accurate ranking of conformer energies. Beyond conformers, the global changes between neighboring geometries generated by time evolution in molecular dynamics or along a reaction path present other potential targets for compression~\cite{Simon2024Improved}. Categorical changes between atom types could be a way to enable thermodynamic integration and alchemical interpolations between ligands~\cite{Huang2026fullqubit}.  

Higher-dimensional potential energy surfaces may offer larger opportunities for compression as the number of geometries grows. At the same time, higher-dimensional grids rapidly increase the cost of classical preprocessing, creating a practical limit on the dimensionality that can be treated directly. One possible way to mitigate this cost is to apply compression to block encodings of the generating functions for Hamiltonian matrix elements rather than to their explicitly computed numerical values. On-the-fly evaluation of Gaussian matrix elements currently carries large prefactors, however, and its crossover with explicit matrix-element encoding occurs at system sizes larger than those considered here~\cite{Babbush2018Exponentially}. An alternative is to formulate the problem in first-quantized electronic-structure algorithms~\cite{su2021fault,da2025comprehensive,Dutkiewicz2026Spectral}, using a space-centered basis in which the geometry-dependent Hamiltonian can be encoded directly from nuclear and electronic coordinates.

Taken together, these results establish SVD-THC as a primitive for quantum-parallel simulation of ensembles of related molecular Hamiltonians. The observed sublinear scaling and large Toffoli reductions show that sharing the representation of many geometries can move ensemble simulation beyond the cost of independent calculations. Further reductions in data-loading and classical preprocessing costs will determine how broadly this advantage can translate into practical quantum algorithms, but the framework provides a route toward using quantum parallelism for molecular problems that require information from many related geometries.

\section*{Acknowledgments}
\label{sec:acknowledgments}
The authors would like to thank Eric Johnston for his invaluable contributions to speeding up the QRE code. We would also like to thank Carter M.~Gustin for showing us that two rounds of amplitude amplification are indeed sufficient for USP circuits in THC,  Modjtaba Shokrian Zini and Bjorn K.~Berntson for their advice on QSP, as well as Michael J.~Bremner and Philipp T.~Ernst for helpful suggestions on the manuscript.
The authors thank Clemens Utschig-Utschig for his comments
on the manuscript and the support during the project.
The authors acknowledge the use of AI in the code development for this project.

\bibliographystyle{apsrev4-2}
\bibliography{references}

\appendix

\section{QPE using quantum signal processing}
\label{sec:newpe}
In this section, we define a quantum phase estimation (QPE) algorithm mixing phase oracles that approximate a time evolution operator differently well. Mixing approximations allows us to combine our SVD-THC circuit with quantum signal processing (QSP) without blowing up the complexity of QPE. The main result of our analysis is that the right recombination of different QSP routines yields a version of QPE that differs very little from an idealized version using queries to the time evolution operator. Indeed, with  $O(n_{\mathrm{p}}\log \varepsilon^{-1})$ additional queries to the SVD-THC circuit, the statistical confidence of an $n_{\mathrm{p}}$-qubit QPE with mixed QSP approximations can be made $(\varepsilon/2)$-close to the ideal QPE. To prove this statement, let us first introduce some notation.

Let 
\begin{align}
e^{\pm i \arccos E_{\theta,k}}\, ,
\end{align}
be the  eigenvalues of the qubitization walk operators featuring the block encoding of   $\hat{H}_{\theta}/\lambda = \sum_k E_{\theta,k} |\psi_{\theta,k}\rangle\!\langle\psi_{\theta,k}|$, 
where the signs `$+$' and `$-$' correspond to different  eigenstates $|Q_{\theta, k, +}\rangle$ and $|Q_{\theta, k, -}\rangle$. Using a Jacobi-Anger expansion polynomial in quantum signal processing \cite{low2019hamiltonian}, these two eigenstates can be approximately recombined into a unitary with eigenvalues $\exp(iE_{\theta,k} t)$ and eigenstates
\begin{align}
|\psi_{\theta,k}\rangle_{\mathsf{s}} |0\rangle_{\mathsf{x}}|0\rangle_{\mathsf{u}}
\end{align}
where $t$ is a real parameter, $\mathsf{u}$ is an auxiliary qubit used in generalized quantum signal processing (GQSP) \cite{motlagh2024generalized}, and $\mathsf{x}$ is the amalgamated auxiliary register for the block encoding $\mathcal{B}[\hat{H}_{\theta}/\lambda]$, containing all permanent (i.e.~non-elbowed) registers of $\mathsf{b}$, $\mathsf{g}$, $\mathsf{j}$, $\mathsf{q}$ and $\mathsf{v}$.  A polynomial of finite degree can unfortunately only implement this time-evolution operator up to some error $\varepsilon$. Suppressing this approximation error is relatively easy, as the degree $d$ of the polynomial, and therefore the number of queries to $\mux_\theta \mathcal{B}[\hat{H}_{\theta}/\lambda]$ in a QSP circuit has with $d=O(t + \log \varepsilon^{-1})$ a rather forgiving scaling with $\varepsilon$. However, we can only benefit from this scaling if we admit different approximations to time evolution to QPE.

For the quantum phase estimation routine in Figure~\ref{fig:QPE}, we define phase-kickback oracles $\mathcal{U}_{j\theta\pm}$ with  GQSP routines depicted on the right-hand side of Figure~\ref{fig:QPE}$(b)$. Every such routine $\mathcal{U}_{j\theta\pm}$ shall have different degrees $d_j$, time parameters $t_j$ and target errors $\varepsilon_j$. Approximating the time evolution operator $\exp(\pm i\hat{H}_{\theta}t_j/\lambda)$ within the all-zero component $\mathcal{A}(\mathcal{U}_{j\theta\pm})$ of the GQSP circuit:
\begin{align}
\mathcal{U}_{j\theta\pm} = \mathcal{A}(\mathcal{U}_{j\theta\pm})\,  |0\rangle\!\langle0|_{\mathsf{u}} |0\rangle\!\langle0|_{\mathsf{x}} + ...\, ,
\end{align}
the approximation error is defined such that
\begin{align}
\label{eq:phase_approx}
    \varepsilon_j \geq \left\| \vphantom{\sum} \mathcal{A}(\mathcal{U}_{j\theta\pm}) -  e^{\pm i\hat{H}_{\theta}t_j/\lambda}\right\|\, .
\end{align}
 This is done for all phase-kickback routines $\mathcal{U}_{j \theta\pm}$ in the full QPE with $n_{\mathrm{p}}$ phase qubits $\mathsf{p}_j$ for $j = 1 \dots n_{\mathrm{p}}$. In this QPE, we unfortunately cannot transfer the approximation errors onto the Hamiltonian, i.e.~we cannot regard the estimated phase as the energy of an effective Hamiltonian shaped by approximation errors. This is due to every phase oracle $\mathcal{U}_{j\theta\pm}$ being an individual approximation of its corresponding time evolution operator $\exp(\pm i \hat{H}_{\theta} t_j/\lambda)$ with
 
\begin{align}
\mathcal{U}_{j\theta\pm} \neq (\mathcal{U}_{1\theta+})^{\pm 2^{j-1}}\, .
\end{align}

In fact, we do not even guarantee a clean unitarity of forwards and backwards evolutions as $\mathcal{U}_{j\theta+} \mathcal{U}_{j\theta-}  = 1$ is fulfilled only approximately. Rather than one being the inverse of the other, we define $\mathcal{U}_{j\theta-}$ as a version of $\mathcal{U}_{j\theta+}$, in which the sign of the Hamiltonian block encoding $\mathcal{B}[\hat{H}_{\theta}/\lambda]$ is flipped,  which makes for an inexpensive double phase-kickback by CZ gates between qubits $\mathsf{p}_j$ and $\mathsf{u}$ as shown in Figure~\ref{fig:QPE}$(b)$. 

So while we cannot hope for a systematic error of the phase, we can make statements about the error of estimated phases. To that end, let us first define the window filter by the coefficients $\lbrace w_r \rbrace$. The window state preparation routine creates the superposition
\begin{align}
    \mathsf{window}\, |0\rangle_{\mathsf{p}} = \sum_r w_r|r\rangle_{\mathsf{p}}
\end{align}
on the phase register $\mathsf{p}$, where integers $r$ are represented as a bit string $r_1 r_2\cdots r_{n_{\mathrm{p}}}$. We now compare the outputs of the QPE in Figure~\ref{fig:QPE} within the $|0\rangle$ subspaces of register $\mathsf{x}$ and qubit $\mathsf{u}$ to a QPE where the phase-kickback routines $\mathcal{U}_{j\theta\pm}$ are perfect time evolution routines $\exp(\pm i\hat{H}_{\theta}t_j/\lambda)$. We want to evaluate the difference
\begin{align}
   \left\|  \mathsf{QFT}\sum_r w_r |r\rangle_{\mathsf{p}} \exp \! \left[ i\hat{H}_{\theta} \sum_{m} (-1)^{r_m}\,  t_m/\lambda \right]|\psi\rangle_{\mathsf{s}} - \mathsf{QFT}\sum_r w_r |r\rangle_{\mathsf{p}} \left[\prod_m \mathcal{A}\!\left(\mathcal{U}_{m\theta(-1)^{r_m}}\right)\right]|\psi\rangle_{\mathsf{s}} \right\|\, .
\end{align}
Disregarding the quantum Fourier transform as a constant unitary, we can define the two states as the beginning and end of a telescoping series of states
\begin{align}
    |\mathsf{QPE}_k\rangle = \sum_{r=0}^{2^{n_{\mathrm{p}}}-1} w_r |r\rangle_{\mathsf{p}}\left[\prod_{j = 1}^{k} e^{i (-1)^{r_j}\hat{H}_{\theta}t_j/\lambda}\right] \left[ \prod_{m=k+1}^{n_{\mathrm{p}}} \mathcal{A}\!\left(\mathcal{U}_{m\theta(-1)^{r_m}}\right)\right]|\psi\rangle_{\mathsf{s}}\, ,
\end{align}
allowing us to evaluate the difference of the states $|\mathsf{QPE}_{n_{\mathrm{p}}}\rangle$ and $|\mathsf{QPE}_0\rangle$ using multiple applications of the triangle inequality:
\begin{align}
    \left\|\vphantom{\sum} |\mathsf{QPE}_0 \rangle - |\mathsf{QPE}_{n_{\mathrm{p}}} \rangle\right\| &\leq \left\|\vphantom{\sum} |\mathsf{QPE}_0 \rangle - |\mathsf{QPE}_1 \rangle\right\| + \left\|\vphantom{\sum} |\mathsf{QPE}_1 \rangle - |\mathsf{QPE}_2 \rangle\right\| + ... +\left\|\vphantom{\sum} |\mathsf{QPE}_{n_{\mathrm{p}}-1} \rangle - |\mathsf{QPE}_{n_{\mathrm{p}}} \rangle\right\| \notag \\
    &= \sum_{k=1}^{n_{\mathrm{p}}} \left\|\vphantom{\sum} |\mathsf{QPE}_{k-1} \rangle - |\mathsf{QPE}_{k} \rangle\right\|\, .\label{eq:qpesegments}
\end{align}

Singular elements of the sum can be evaluated as
\begin{align}
    &\left\|\vphantom{\sum} |\mathsf{QPE}_{k-1} \rangle - |\mathsf{QPE}_{k} \rangle\right\| \notag \\ 
    &  =  \notag \left\| \sum_{r=0}^{2^{n_{\mathrm{p}}}-1} w_r |r\rangle_{\mathsf{p}}\left[\prod_{j = 1}^{k-1} e^{i (-1)^{r_j}\hat{H}_{\theta}t_j/\lambda}\right] \left[\vphantom{\prod_k^n}   \mathcal{A}\!\left(\mathcal{U}_{k\theta(-1)^{r_k}}\right) - e^{i (-1)^{r_k}\hat{H}_{\theta}t_k/\lambda}\right]\right.\\
    & \quad\quad\times\left.\left[ \prod_{m=k+1}^{n_{\mathrm{p}}} 
    \mathcal{A}\!\left(\mathcal{U}_{m\theta(-1)^{r_m}}\right)\right]|\psi\rangle_{\mathsf{s}} \right\| \\ 
    & \leq  \sqrt{\sum_r |w_r|^2 \left\| \left[\vphantom{\prod_k^{n_{\mathrm{p}}}}   \mathcal{A}\!\left(\mathcal{U}_{k\theta(-1)^{r_k}}\right) - e^{i (-1)^{r_k}\hat{H}
    _{\theta}t_k/\lambda}\right]\left[ \prod_{m=k+1}^{n_{\mathrm{p}}} \mathcal{A}\!\left(\mathcal{U}_{m\theta(-1)^{r_m}}\right)\right]|\psi\rangle_{\mathsf{s}}\right\|^2}\\
    \label{eq:oneqpesegment_0}
     & \leq \sqrt{\sum_r |w_r|^2 \left\| \mathcal{A}\!\left(\mathcal{U}_{k\theta(-1)^{r_k}}\right) - e^{i (-1)^{r_k}\hat{H}_{\theta}t_k/\lambda}  \right\|^2} \\
     & \leq   \underbrace{\sqrt{\sum_r |w_r|^2}}_{=1} \; \cdot \;  \varepsilon_k , \label{eq:oneqpesegment}
\end{align}
where first we have used $\|\sum w_r |r\rangle \mathsf{A}_r|\Psi_r\rangle \|\leq \sqrt{\sum_r |w_r|^2 \| \mathsf{A}_r|\Psi_r\rangle\|^2}$ for any operators $\mathsf{A}_r$ and states $|\Psi_r\rangle$, then we have used the fact that operator norm is defined as the supremum of all supporting states, $\|\mathsf{A}\| = \sup_{|\Psi\rangle} \| \mathsf{A}|\Psi\rangle \|$, including the choice $|\Psi\rangle = \left[\prod_{m>k} \mathcal{A}\!\left(\mathcal{U}_{m\theta(-1)^{r_m}}\right)\right]|\psi\rangle$. Finally, \eqref{eq:oneqpesegment_0} is recombined with \eqref{eq:phase_approx}. Plugging \eqref{eq:oneqpesegment} into \eqref{eq:qpesegments}, we find the difference between states output by ideal and approximate QPE to be $\sum_k\varepsilon_k \leq \varepsilon$, following the definition of $\varepsilon_k = \varepsilon 2^{-(n_{\mathrm{p}}+1-k)}$ in Section~\ref{sec:threshold}. This result might directly influence the distribution of phase values, i.e., if the output states are
\begin{align}
    |p\rangle = \sum_{j=0}^{2^{n_{\mathrm{p}}}-1} \sqrt{p_j}|j\rangle_{\mathsf{p}}|\chi_j\rangle_{\mathsf{sux}}\quad \text{and} \quad 
    |\widetilde{p}\rangle = \sum_{j=0}^{2^{n_{\mathrm{p}}}-1} \sqrt{\widetilde{p}_j}|j\rangle_{\mathsf{p}}|\chi_j\rangle_{\mathsf{sux}}\, ,
\end{align}
for some states $|\chi_j\rangle$ and with  $\sum_j |p_j-\widetilde{p}_j| \leq \varepsilon$. In the worst case, we lose $\varepsilon / 2$ of squared amplitude inside the confidence region of QPE, and gain  $\varepsilon/2$ in probability of measuring the phase in the outside region of our confidence interval. For instance, if the only phase $\alpha$ measured is in between  $2\pi y 2^{-n_{\mathrm{p}}}$ and $2\pi (y+1)2^{-n_{\mathrm{p}}}$, then $p_y + p_{y+1}-\widetilde{p}_y - \widetilde{p}_{y+1}  \leq \varepsilon/2$. That is, we are just lowering the probability of measuring phases with error $\pi 2^{-n+1}$ by $\varepsilon / 2$.

The alternative, in which we do not mix approximations to the time evolution, would have been much less favorable. Using QSP to define a constant phase oracle $U$ that is  $\widetilde{\varepsilon
}$-close to $\exp(i\hat{H}_{\theta}t_1/\lambda)$ would cause an energy error of size $O(\lambda \widetilde{\varepsilon})$ while increasing the complexity of QPE by a factor $O(\log \widetilde{\varepsilon}^{-1})$.

\section{Inner state preparation for SVD-THC}
\label{sec:prepare}
\begin{figure}
    \centering
    \includegraphics[width=0.7\linewidth]{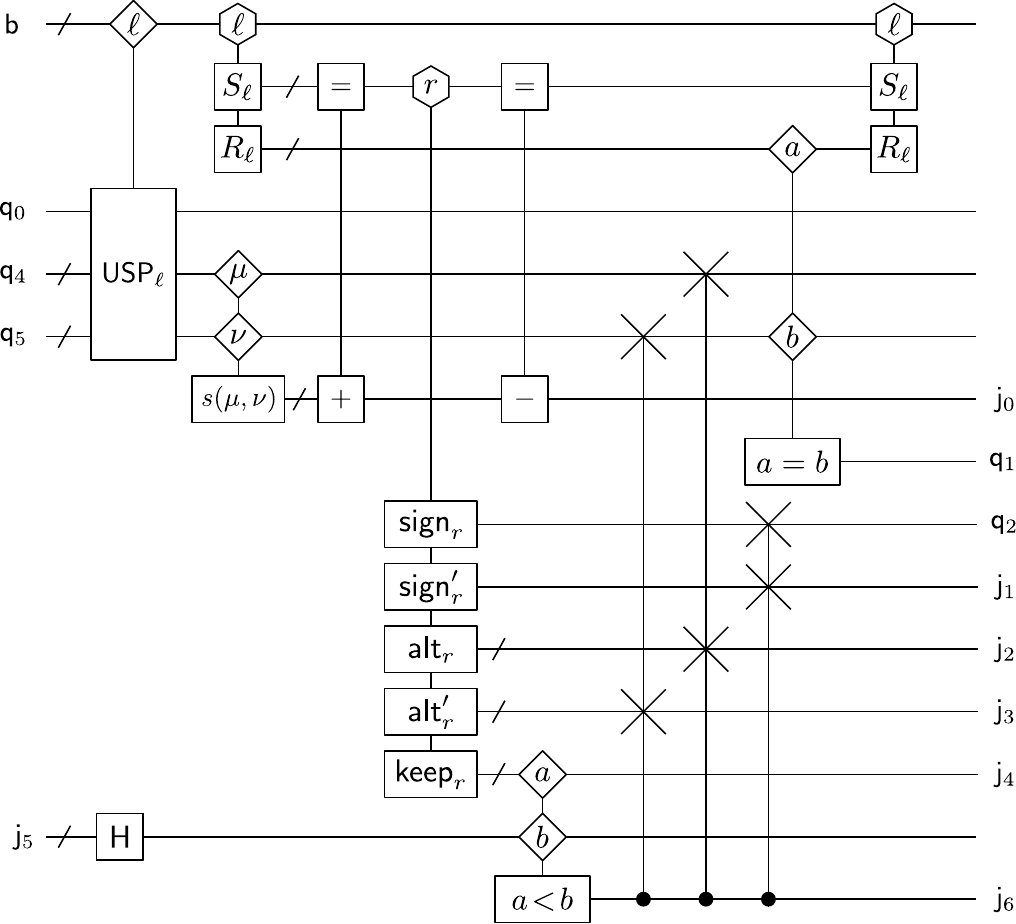}
    \caption{Circuit construction for $\mux_{\ell}\mathsf{prep}_{\mathsf{i},\ell}$ of SVD-THC. On the basis of a multiplexed uniform state preparation circuit $(\mux_\ell \mathsf{USP}_\ell)$, this circuit encodes the state $\propto \sum_{\mu\nu}\zeta_{\ell\mu\nu }|\mu\rangle|\nu\rangle$ with coefficients $\zeta_{\ell\mu\nu}$, triple-indexed by the values $\mu$ and $\nu$ in registers $\mathsf{q}_4$ and $\mathsf{q}_5$, as well as the superindex $\ell$ in $\mathsf{b}$. As a first step, the indices $\mu$ and $\nu$ are turned into a continuous index $s(\mu,\nu) = \nu (\nu + 1)/2  + \mu$ with quantum arithmetic circuits. Since the indices $\mu$ and $\nu$ run in the ranges $[0,R_\ell)$ and $[0, R_\ell]$, upper-limited by the $\ell$-dependent THC ranks $R_\ell$, we load the bases $S_\ell$ for the indices $r(\ell,\mu,\nu)$ into a temporary register. An in-place addition with the $s(\mu,\nu)$ values then completes the computation of the index $r$, from which we can load $\mathsf{alt}_r$ and $\mathsf{alt}_r^\prime$, the alternative values for $\mu$ and $\nu$ as well as the keep probabilities $\mathsf{keep}_r$, along with original and alternative values for the signs, $\mathsf{sign}_r$ and $\mathsf{sign}^\prime_r$. A comparator -- the gate labeled `$a<b$' -- between the keep probabilities and the values in register $\mathsf{j}_5$ controls the swaps of original and alternative values. To flag the scenario of  $\nu=R_\ell$ in the  $|\ell\rangle$ subspace of $\mathsf{b}$, we load the THC ranks $R_\ell$ with $S_\ell$, and set the qubit $\mathsf{q}_1$ with what is largely a multi-Toffoli gate, labeled `$a=b$'. To uncompute temporary registers, the in-place addition of $r$ is reverted.} 
    \label{fig:mux_prep}
\end{figure}

The multiplexed version of the inner state preparation, $\mathsf{prep}_{\mathsf{i},\ell}$ utilizes a modified version of Alias Sampling to prepare auxiliary state \eqref{eq:prepstate}.

The circuit presented in Fig.~\ref{fig:mux_prep} features a data-loader with aliases $\mathsf{alt}$ and $\mathsf{alt^\prime}$ for indices $|\mu\rangle$ and $|\nu\rangle$ respectively, while loading also a sign into $\mathsf{q}_2$ as well as an alternative sign into the junk register $\mathsf{j}_1$. All these quantities are triply indexed, with $\mu\leq\nu$, and in the $|\ell\rangle$ subspace of $\mathsf{b}$, the indices  $(\mu, \nu)$ assume all possible values from $(0,0)$ to $(N-1, R_\ell)$.  This means that the size of registers $\mathsf{q}_4$ and $\mathsf{q}_5$ must be padded to the largest possible THC rank, bringing their size to $\lceil \log(\max_\ell R_\ell + 1) \rceil$ qubits. Since $|\mu\rangle$ and $|\nu\rangle$ are swapped later in $\mathsf{Select}$, $\mathsf{q}_4$ and $\mathsf{q}_5$ must have the same size. In the state-of-the-art THC circuit \cite{caesura2025faster}, the indices $\mu$ and $\nu$ are joined arithmetically into $s(\mu, \nu) = (\nu+1)\nu/2 + \mu$. This contiguizer would also work here, where the indices are padded. We obtain the triply joined index $r(\ell,\mu,\nu)$ by adding the value of $s$ to $S_\ell$, the basis with respect to the most significant index $\ell$:

\begin{align}
    r(\ell,\mu,\nu) = S_\ell + s(\mu,\nu)\, 
\end{align}

where 

 \begin{align}
     S_\ell = \ell N + \sum_{m<\ell} \frac{(R_m+1)R_m}{2}\, .
 \end{align}
As part of the circuit, we also set qubit $\mathsf{q}_1$ in the subspace $|\ell\rangle|\nu=R_\ell\rangle$ of qubits $(\mathsf{b},\mathsf{q}_5)$, as a flag for single-particle Hamiltonians. To this end, we load the value of $R_\ell$ in the very same routine loading  $S_\ell$. After the controlled swaps of the circuit, the value can then be XORed into $\mathsf{q}_5$, and $\mathsf{q}_1$ is set by a multi-Toffoli gate.

\begin{sidewaysfigure}
    \centering
\includegraphics[width=\linewidth]{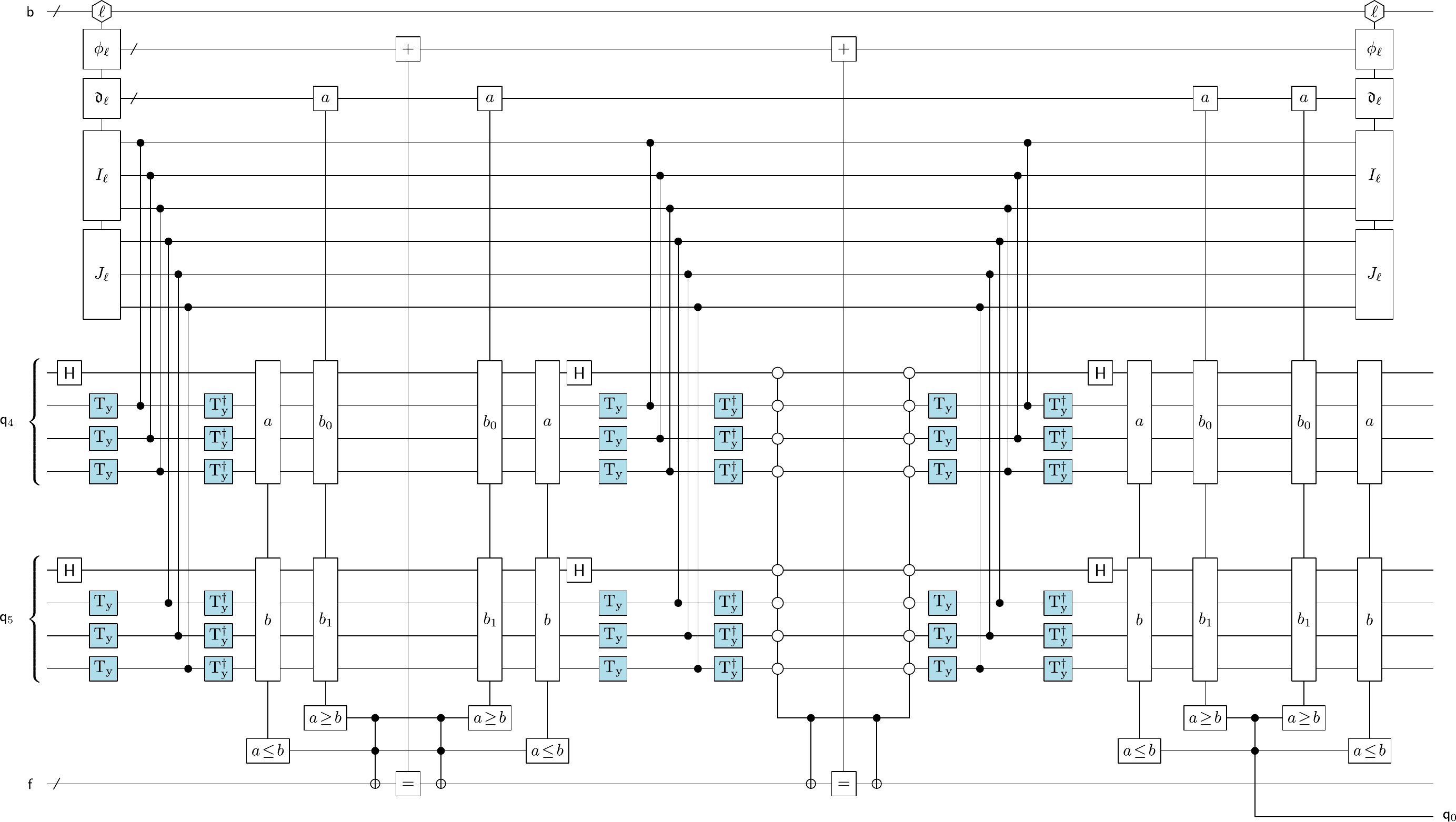}
    \caption{The $\mux_\ell \mathsf{USP}_\ell$ routine of the multiplexed THC block encoding, preparing the registers $\mathsf{q}_4$ and $\mathsf{q}_5$ in the uniform state $\propto\sum_{\mu\nu}|\mu\rangle|\nu\rangle$ with $\mu\leq \nu$ and indices in the range from $(\mu, \nu) = (0,0) ... (N-1, R_\ell)$, where the THC ranks $R_\ell$ depend on the superindex $\ell$ in register $\mathsf{b}$. In this toy example of $\mathsf{USP}_\ell$,  $\mathsf{q}_4$ and $\mathsf{q}_5$ contain 4 qubits each, and the uniform state is prepared within a single round of (fractional) amplitude amplification. Note that a realistic circuit might require two rounds of amplitude amplification. The $\ell$-dependent angles $\phi_\ell$ for partial reflections, dimensional integers $\mathfrak{d}_\ell$ as well as bit masks $I_\ell$ and $J_\ell$ are loaded into temporary registers with comparators that set the bit under the gates `$a\leq b$' and  `$a\geq b$'. The state of this check bit controls a flip (gate `$\oplus$') of the entire register $\mathsf{f}$, holding a phase gradient state, that effects an inversion of the phase resulting from the subsequent in-place addition of the angles $\phi_\ell$ into $\mathsf{f}$. Due to the  $\ell$-dependence of $R_\ell$, we prepare a different number of qubits in $\mathsf{q}_4$ and $\mathsf{q}_5$ in $|+\rangle$  by using the bit masks $I_\ell$ and $J_\ell$ as stencils, where we implement controlled Hadamard gates with CZ gates and T gates in the $Y$ basis, $T_{\mathrm{y}}$, as well as their inverses. After tagging the initial state using Toffoli gates, we set the qubit $\mathsf{q}_0$ in the good subspace, using the same checks as before.}
    \label{fig:mux_USP}
\end{sidewaysfigure}

A resource state for Alias Sampling is prepared approximately by a multiplexed version of the subroutine $\mathsf{USP}_\ell$. We find

\begin{align}
\mathsf{USP}_\ell|0\rangle_{\mathsf{q}_0\mathsf{q}_4\mathsf{q}_5} =  |1\rangle_{\mathsf{q}_0}\sum_{\mu,\nu} \Lambda_{\ell\mu\nu}|\mu\rangle_{\mathsf{q}_4}|\nu\rangle_{\mathsf{q}_5}
\end{align}

with uniform but constrained coefficients 

\begin{align}
\label{eq:constrains}
\Lambda_{\ell\mu\nu} = \begin{cases}
    \sqrt{\frac{1}{(R_\ell+1)R_\ell/2 + N}} & \mu\leq \nu\;\;  \text{and} \;\; \nu\oplus\mu \leq R_\ell\oplus (N-1)\\
    \\
    0& \text{else,}
\end{cases} 
\end{align}

and where qubit $\mathsf{q}_0$ certifies success. In regular THC, the USP circuit would start from a uniform superposition $\propto\sum_{\mu,\nu}|\mu\rangle|\nu\rangle$ of all possible values of $\mu$ and $\nu$, which can then be constrained with at most two rounds of amplitude amplification, depending on the THC rank. In the multiplexed $\mathsf{USP}_\ell$ circuit, depicted in Figure~\ref{fig:mux_USP}, registers $\mathsf{q}_4$ and $\mathsf{q_5}$ are padded up to the largest THC rank $\max_\ell R_\ell$. For subspaces $|\ell\rangle$ with dramatically lower ranks $R_\ell$, this could mean more than two rounds of amplitude amplification. In the circuit, we therefore load bit masks $I_\ell$ and $J_\ell$, that would control Hadamard gates in registers $\mathsf{q}_4$ and $\mathsf{q}_5$, respectively.  CZ gates running transversely from the temporary registers holding $I_\ell$ ($J_\ell$) into a $\mathsf{q}_4$ ($\mathsf{q}_5$), right in between a T gate in $Y$ basis $\mathrm{T_y} = \exp(i\pi/8 Y)$ and its inverse $\mathrm{T^\dagger_y}$ (colored blue in Figure~\ref{fig:mux_USP}), affect the target qubit as

\begin{align}
    \mathrm{T^\dagger_y}\, Z\, \mathrm{T^{\phantom{\dagger}}_y}  = \mathsf{H}\, ,
\end{align}

when the respective qubit in the bit mask was set to $|1\rangle$. This circuit restricts how many qubits in $\mathsf{q}_4$ and $\mathsf{q}_5$ are prepared in $|+\rangle$. The result is a precursor state, from which the target superposition state can be reached with two rounds of amplitude amplification in every subspace $|\ell\rangle$. As the circuit in Fig.~\ref{fig:mux_USP} demonstrates, the registers $\mathsf{q}_4$ and $\mathsf{q_5}$ may have a static part of $\lceil\log(\min_\ell R_\ell +1) \rceil$ qubits, where free Hadamard gates suffice as a bit mask would never switch them off. The constrained subspace $|\Lambda_\ell\rangle =\sum_{\mu,\nu}\Lambda_{\ell\mu\nu}|\mu\rangle|\nu\rangle$ is then tagged with quantum arithmetic circuits. From \eqref{eq:constrains}  we can read out the two constraints for $|\mu\rangle|\nu\rangle$:

\begin{enumerate}
    \item the index on $\mathsf{q}_4$ is always lower than or equal to the index in $\mathsf{q}_5$; $\mu\leq \nu$; and
    \item the number $\nu\oplus \mu$, derived from appending the binary representations of $\mu$ and $\nu$ has an upper bound $\nu\oplus \mu\leq R_\ell\oplus (N-1)$.
\end{enumerate}

Both constraints are tested with comparator circuits and the space in which they are mutually fulfilled is flagged. For the second constraint, the $\ell$-dependent upper bound $\mathfrak{d}_\ell$ has been loaded into a temporary register. Some $\ell$-dependent angles are loaded in as well: they are parameters for partial reflections that prevent overcooking the state in the amplification process. Bit masks, $\mathfrak{d}_\ell$  parameters and angles may be loaded in at the beginning of the circuit, and uncomputed as it terminates. While the ideal state $|\Lambda_\ell\rangle$ can be prepared exponentially well for all $\ell$, we nonetheless elect to set the bit $\mathsf{q}_0$ in the $|\Lambda_\ell\rangle$ subspace of the resulting state, by re-using the tagging circuits from the reflections on $|\Lambda_\ell\rangle$.

\section{Comparison against sequential QPE}
\label{sec:sequential}

In this section, we compare the proposed SVD-THC methodology against the simplest alternative without compression,  which we refer to as sequential QPE. In this approach, an ordinary single-geometry phase estimation is performed for each of the $M$ geometries. Each QPE has an individual rank and 1-norm, and quantum resource estimates depend on the average 1-norm $\bar\lambda$ and average rank $\bar R$, as well as length $\bar L = N + \langle R(R+1)/2 \rangle$. This approach is somewhat different from the sliced-THC thresholding circuit of Sec.~\ref{sec:qre}, which slices BLISS-THC circuits for all $M$ geometries coherently, where each Hamiltonian is normalized with the worst-case 1-norm $\lambda_{\max}$ so the circuit is valid at every geometry. As highlighted in Table~\ref{tab:naive-baseline-comparison},   sequential QPE, which could very well be used as an alternative to a thresholding circuit, beats a sliced-THC version of the circuit by factors of $3.3\times$ to $3.8\times$ in Toffoli gates. Against sequential QPE, the full-grid Toffoli improvement of the SVD-THC version of the thresholding circuit using the gate-optimal setting is $11.8\times$ and $10.1\times$ for rigid and relaxed BCE, $38.2\times$ for glycine dipeptide and $60.2\times$ for alanine dipeptide, compared to the respective improvement factors of $41.4\times$, $34.9\times$, $144.5\times$ and $200.6\times$ on the sliced-THC version of the algorithm.
\begin{table*}[t]
\centering
\small
\setlength{\tabcolsep}{2.5pt}
\setlength{\aboverulesep}{0pt}\setlength{\belowrulesep}{0pt}
\renewcommand{\arraystretch}{1.3}
\begin{tabular}{lccc|ccc|cccc}
\toprule
& & & & \multicolumn{3}{c|}{Sequential QPE} & \multicolumn{4}{c}{Thresholding circuit} \\
& & & & \multicolumn{3}{c|}{{BLISS-THC}} & \multicolumn{4}{c}{{Sliced-THC}} \\
\cmidrule{5-11}
System & $M$ & $\bar R$ & $\bar L$ & $\bar\lambda$ & AV & Toff. & $\lambda_{\max}$ & AV & Toff. & Ratio \\[-2pt]
& & & & {\footnotesize(Ha)} & & & {\footnotesize(Ha)} & & & \\
\midrule
BCE rigid & 73 & 51.5 & 1{,}379 & 35.79 & $1.82\!\times\!10^{12}$ & $2.38\!\times\!10^{10}$ & 41.88 & $6.95\!\times\!10^{12}$ & $8.35\!\times\!10^{10}$ & $\mathbf{3.5\times}$ \\
BCE relaxed & 73 & 51.6 & 1{,}386 & 36.45 & $1.85\!\times\!10^{12}$ & $2.42\!\times\!10^{10}$ & 50.45 & $7.14\!\times\!10^{12}$ & $8.36\!\times\!10^{10}$ & $\mathbf{3.5\times}$ \\
Glycine dipep. & 2{,}664 & 134.3 & 9{,}172 & 131.69 & $8.35\!\times\!10^{14}$ & $7.46\!\times\!10^{12}$ & 134.95 & $3.25\!\times\!10^{15}$ & $2.82\!\times\!10^{13}$ & $\mathbf{3.8\times}$ \\
Alanine dipep. & 5{,}184 & 148.8 & 11{,}249 & 149.96 & $2.17\!\times\!10^{15}$ & $1.85\!\times\!10^{13}$ & 153.68 & $7.72\!\times\!10^{15}$ & $6.15\!\times\!10^{13}$ & $\mathbf{3.3\times}$ \\
\bottomrule
\end{tabular}
\caption{The two  references for the thresholding circuit without compression. Sequential QPE sums $M$ independent single-geometry phase estimations at the average rank $\bar R$ and 1-norm $\bar\lambda$ and average length $\bar L=N+\langle R(R+1)/2\rangle$. The thresholding circuit with sliced-THC  contains the same $M$ geometry-conditioned BLISS-THC circuits and schedules its Jacobi-Anger expansion at the sampled worst-case 1-norm $\lambda_{\max}$ while reusing  workspace. `$\mathrm{Ratio}$' is the relative improvement of sequential QPE over the sliced-THC thresholding circuit in terms of Toffoli gates. Both references are validated with CCSD, as SVD-THC is: each per-geometry factorization is fitted without correlation-energy reweighting and accepted at the smallest THC rank whose CCSD correlation-energy error is below $0.6$~mHa.}
\label{tab:naive-baseline-comparison}
\end{table*}

We attribute this difference to how the two references are counted rather than to the algorithms themselves. First, the thresholding circuit of Fig.~\ref{fig:thresh} performs a phase estimation and its inverse in every iteration, and both multiplexed approaches are charged for the two, whereas sequential energy estimation requires only one. Second, the Jacobi-Anger schedule rounds $\lambda/\epsilon$ up to the next power of two, while the sequential reference uses the integer estimate $\lceil \pi\bar\lambda/\epsilon \rceil$. At $\epsilon=1$~mHa this rounding costs a factor of $1.83$ and $1.80$ for rigid and relaxed BCE, $1.99$ for glycine dipeptide, whose $\bar\lambda/\epsilon$ lies only $0.5\%$ above a power of two, and $1.75$ for alanine dipeptide. The remaining $4$--$5\%$ reflects a small difference in the per-query cost of the two estimates. We note that the worst-case norm, which exceeds $\bar\lambda$ by $2.5$--$38\%$, falls on the same power of two as $\bar\lambda$ in every system and therefore does not increase the query count, and that both references use the same phase-estimation normalization. When the two references are counted consistently, the improvement of SVD-THC takes the closed form,
\begin{align}
    \mathcal{I} = M\;\frac{c_{\mathrm{sliced}}}{c_{\mathrm{SVD}}}\;\frac{\mathcal{N}_{\mathrm{sliced}}}{\mathcal{N}_{\mathrm{SVD}}}\, ,
    \label{eq:improvement_closed_form}
\end{align}
where $c_{\mathrm{x}}$ denotes the cost of a single query and $\mathcal{N}_{\mathrm{x}}$ the Jacobi-Anger query count of one quantum circuit. Equation~\eqref{eq:improvement_closed_form} is independent of the number of phase estimations per iteration, applies equally to the sequential and multiplexed references, and reproduces every full-grid improvement of Sec.~\ref{sec:qre}. For glycine dipeptide, for example, $\mathcal{I}=2{,}664\times(6{,}420/59{,}212)\times(1/2)\approx144$, where the factor of one half arises because the multiplexed 1-norm, $2.5$ times the worst-case per-geometry value, places SVD-THC one step higher on the Jacobi-Anger schedule. 

These results suggest that the choice of reference algorithm does not change the underlying advantage of SVD-THC, which is set by the number of geometries and by the relative cost of a compressed query: $9.2\times$ and $12.9\times$  a single-geometry query in Toffolis for glycine and alanine dipeptides, and $76\times$ and $118\times$  in active volume, which explains why the active-volume advantage is an order of magnitude smaller. The two references differ genuinely in two respects. Their qubit counts differ slightly and in either direction, $884$ for sequential QPE against $813$ for the thresholding circuit for glycine dipeptide, and $1139$ against $1191$ for alanine dipeptide. Sequential QPE also returns $M$ classical energies rather than coherent access to the ensemble. For plain energy estimation the sequential approach is therefore sufficient, whereas the thresholding and search applications that motivate this work require coherent access. Because the Jacobi-Anger schedule advances in powers of two, we also recommend that any improvement factor be reported together with the schedule and the number of phase estimations on which it is based.

\end{document}